\documentclass[twocolumn]{aastex631}
\usepackage{amssymb}
\usepackage{amsmath}
\usepackage{bm}
\usepackage{float}
\usepackage{savesym}
\savesymbol{tablenum}
\usepackage{siunitx}
\usepackage{ulem}
\usepackage{xcolor}
\usepackage{soul}
\usepackage{booktabs}
\restoresymbol{SIX}{tablenum}

\definecolor{webbrown}{rgb}{.6,0,0}

\begin{document}

\title{A direct detection of neutral hydrogen intensity mapping on Mpc scales at $z\approx 0.32$ and $z\approx 0.44$}

\author[0000-0002-8671-2177]{Sourabh Paul}
\affiliation{Jodrell Bank Centre for Astrophysics, School of Physics and Astronomy, The University of Manchester, Manchester M13 9PL, UK}
\affiliation{Department of Physics and Astronomy, University of the Western Cape, Robert Sobukhwe Road, Bellville, 7535, South Africa}
\affiliation{Department of Physics, McGill University, Montreal, QC, Canada H3A 2T8}
\thanks{sourabh.paul@gmail.com}

\author[0000-0002-4965-8239]{Zhaoting Chen}
\affiliation{Institute for Astronomy, The University of Edinburgh, Royal Observatory, Edinburgh EH9 3HJ, UK}
\affiliation{Jodrell Bank Centre for Astrophysics, School of Physics and Astronomy, The University of Manchester, Manchester M13 9PL, UK}
\thanks{zhaoting.chen@manchester.ac.uk}

\author[0000-0003-3892-3073]{Mario G. Santos}
\affiliation{Department of Physics and Astronomy, University of the Western Cape, Robert Sobukhwe Road, Bellville, 7535, South Africa}
\affiliation{South African Radio Observatory (SARAO), 2 Fir Street, Observatory, Cape Town, 7925, South Africa}
\thanks{mgrsantos@uwc.ac.za}

\author[0000-0003-3334-3037]{Laura Wolz}
\affiliation{Jodrell Bank Centre for Astrophysics, School of Physics and Astronomy, The University of Manchester, Manchester M13 9PL, UK}

\begin{abstract}
We report the detection of the cosmological power spectrum using the intensity mapping signal from 21-cm emission of neutral hydrogen (HI), derived from interferometric observations with the L-band receivers of the MeerKAT radio telescope. Intensity mapping is a promising technique to map the three-dimensional matter distribution of the Universe at radio frequencies and probe the underlying Cosmology. So far, detections have only been achieved through cross-correlations with galaxy surveys. Here we present independent measurements of the HI power spectrum at redshifts $0.32$ and $0.44$ with the foreground avoidance method. We utilize two distinct frameworks for mitigating systematics, where a conservative baseline flagging based approach achieves detections at $3.2\sigma$ and $3.5\sigma$, and a power spectrum based flagging method enhances the significance to $5.9\sigma$ and $9.18\sigma$, respectively. The information contained in the power spectrum measurements allows us to probe the parameters of the HI mass function and HI halo model. These results are a significant step towards precision cosmology with HI intensity mapping using the new generation of radio telescopes.

\end{abstract}


\section{Introduction} \label{sec:intro}
The 21-cm hyperfine transition of neutral hydrogen (HI) is a prime tracer of the HI gas in galaxies as well as the inter-galactic medium at high redshifts. Observations at the radio frequencies of this line (at or below $1420$ MHz) are free from dust absorption and have high redshift resolutions due to standard digital processing techniques from radio telescopes. Given the abundance of HI gas in the Universe, observations with the 21-cm line can be a powerful probe of the structure and evolution of the Universe. However, the inherent weakness of the 21-cm signal has restricted HI galaxy surveys to the nearby Universe (redshifts $\lesssim 0.1$; e.g. \citet{Jones_2018}). The upcoming generation of radio telescopes will push this to higher redshifts but only over small areas and large HI masses \citep{Maddox_2021}. The intensity mapping (IM) technique circumvents this sensitivity issue by averaging over the collective emission from many unresolved galaxies and producing low angular resolution maps of the HI intensity (albeit still with redshift resolutions comparable to galaxy spectroscopic surveys). The possibilities
opened by this new observational window promise a breakthrough in constraining cosmological theories in the very near future \citep{Bharadwaj_Sethi_2001, 2004MNRAS.355.1339B,2010ARA&A..48..127M, Bull_2015}. 

However, the continuous mapping of the skies requires separating the 21-cm line from other signals at a given frequency.
This is particularly challenging due to the presence of strong radio foregrounds \citep{Wolz_2014, Olivari_2016, Cunnington_2019, Liu_2020}. 
So far, measurements have been hampered by the extraordinary calibration challenges and detections have only been achieved through cross-correlations with galaxy surveys since the systematics in the  HI maps are not expected to correlate with data at other wavelengths. The first cross-correlation detection was obtained with the single-dish Green Bank Telescope (GBT)\citep{Chang_2010,Masui_2013,Wolz_2022}, followed by measurements with the Parkes telescope \citep{Anderson_2018}. More recently, the CHIME interferometer made a stacking detection using this technique over a wide area \citep{CHIME_2022} (but see also \citet{Chowdhury_2020}).

The MeerKAT telescope array in South Africa has also been proposed as an exquisite survey machine for HI IM \citep{Santos_2017} and acts as a precursor for the upcoming SKA Observatory \citep{Santos_2015,SKA_2018}.
The telescope has a 64-dish antenna configuration, each antenna having a diameter of $13.5$ m. The core of the array consists of 48 antennas within 1 km of diameter. The rest of the antennas are spread over a radius of up to 4 km, which provides a higher angular resolution. The dense core distribution results in a large number of short baselines, thereby increasing the sensitivity in low $k_\perp$ modes, i.e. the larger cosmological scales. With its substantial collecting area and a multitude of baselines, MeerKAT is an exceptionally sensitive instrument that has already achieved successful stacking results \citep{An_2021,Sinigaglia_2022}.

For IM surveys, the array is used in single dish mode in order to access the large cosmological scales and pilot tests have been underway \citep{Wang_2021}. A detection of the HI power spectrum in cross-correlation with galaxies from the WiggleZ Dark Energy Survey has been made \citep{2023MNRAS.518.6262C}. However, the large primary beam limits the minimum accessible scales to tens of Mpc and interferometric IM observations can provide crucial information on smaller scales and the nonlinear regime. The core heavy dish distribution of the MeerKAT interferometer can measure scales up to $60$ Mpc at $z<0.65$ in the \textit{L}-band frequency range (856 MHz $< \nu <$ 1712 MHz) and up to $165$ Mpc at $z<1.4$ in the \textit{UHF}-band (580 MHz $< \nu <$ 1015 MHz). 

{Forecasts in \citet{Paul_2021} predict that a statistical detection of the HI IM power spectrum is possible with $\sim 100$ hrs of observation with MeerKAT. Using surveys such as the MeerKAT International GHz Tiered Extragalactic Exploration (MIGHTEE) Survey \citep{MIGHTEE}, we can detect the HI power spectrum in narrow redshift bins across a range of redshifts, opening up a new window for probing the astrophysics of the HI galaxies \citep{Chen_2021}.}

In this work, we report the first detection of the HI intensity auto-power spectrum in the interferometric IM mode at frequencies of $986$ MHz ($z \approx 0.44$) and $1077.5$ MHz ($z \approx 0.32$) by analysing $\sim 96$ hrs of MeerKAT data. Observations were carried out with the $L$-band configuration and 4,000 frequency channels with a resolution of $209$ KHz, during the commissioning process of MeerKAT in 2018. We estimate the power spectrum from two RFI-free frequency ranges centered at $986$ and $1077.5$ MHz with 220 frequency channels each, equalling a frequency width of $\approx 46$ MHz.

We use the flat $\Lambda$CDM cosmological parameters $[\Omega_m, \Omega_b, h, n_s, \sigma_8] = [0.311, 0.049, 0.677, 0.967, 0.8102]$ from Planck 2018 results \citep{Planck_2018} throughout this work.

\section{MeerKAT data}
\label{Data_calibration}
The observed field J2000, located at $\alpha=04^{\rm{h}}13^{\rm{m}}26.4^{\rm{s}}, \delta=\ang{-80;00;00}$, contains a limited amount of bright radio foreground sources. It was previously studied to produce the deepest radio source count to date using the P(D) technique, revealing the bulk of the star formation history of the Universe \citep{Mauch_2020, Matthews_2021}. This field has been selected for the analysis presented in this paper due to the availability of long integration hours and the absence of bright continuum sources. The observations were carried out with the $L$-band configuration and the data used in this work is a subset of the publicly available data observed under the Proposal ID: SCI-20180426-TM-01 \citep{Mauch_2020}. The 96 hours of data were collected across 9 datasets between June 30, 2018, and November 4, 2018, with each dataset utilizing a minimum of 58 antennas (details in \autoref{table:obs_details}). 
\begin{table}
\centering
\begin{tabular}{| c | c c c c|} 
 \hline
  & & Total & Target & \\
 Obs id &Start Date \& time&duration&duration&N$_{\rm{ant}}$\\
  & (UTC) & (hr) & (hr) & \\
 \hline
 1530399641 & 2018 Jun 30, 23:00 & 16.2 & 12.6 & 60\\
 1530999556 & 2018 Jul 7, 21:39 & 17.2 & 13.4 & 61\\
 1531777026 & 2018 Jul 16, 21:37 & 8.0 & 6.0 & 61\\
 1532466465 & 2018 Jul 24, 21:07 & 8.9 & 6.9 & 59\\
 1532552470 & 2018 Jul 25, 21:01 & 9.0 & 7.6 & 58\\
 1532725253 & 2018 Jul 27, 21:01 & 16.1 & 14.0 & 61\\
 1532811076 & 2018 Jul 28, 20.51 & 16.2 & 14.1 & 60\\
 1538775215 & 2018 Oct 8, 21:33 & 9.5 & 8.5 & 59\\
 1541342249 & 2018 Nov 4, 14:37 & 16.2 & 14.2 & 62\\     
 \hline
\end{tabular}
\caption{Observation summary of the 9 datasets used in this paper. }
\label{table:obs_details}
\end{table}
Each scan on the target field lasted 15 minutes followed by a 2 minutes observation of the secondary (e.g. phase) calibrator $\rm{PKS}\, \rm{J}0252{\text -}7104$. The primary flux and bandpass calibrator $\rm{PKS}\, \rm{B}1934{\text -}638$ was observed for 10 minutes at the beginning of each session and after that, in regular intervals of 3 hours. 

\subsection{RFI flagging and calibration}
The flagging of unwanted radio frequency interference (RFI) from terrestrial sources and subsequent calibration of the data were performed with the \textsc{processMeerKAT} \citep{processmeerkat} software which is built for calibrating MeerKAT interferometer data. The processMeerKAT pipeline uses the Common Astronomy Software Applications (CASA, \citealt{CASA}) package routines for RFI removal, and calculations of phase and flux gains from the reference calibrator observations. The following steps are performed by \textsc{processMeerKAT} on each of the 9 datasets:

The data undergoes an initial processing step with its original time resolution of 8 seconds and is reduced to a sub-band of $952–1170$ MHz. During this stage, the first round of RFI flagging takes place. Specifically, frequency ranges corresponding to persistent RFI bands at $933-960$ MHz and $1163-1310$ MHz are flagged. Additionally, the data is clipped at a threshold of $50$ Jy to remove the strongest RFI signals. The \textsc{tfcrop} algorithm is then applied independently to the primary and secondary calibrators as well as the target using a conservative flagging threshold. Subsequently, the standard delay, bandpass, and gain calibration procedures are carried out on the data. For the estimation of delay and bandpass solutions, the $\rm{PKS}\, \rm{B}1934-638$ field is utilized, and these solutions are applied to the $\rm{PKS}\, \rm{J}0252-7104$ data to calculate the time-dependent complex gains. The final gain corrections are then applied to the target data. In the second stage of RFI flagging, both the \textsc{tfcrop} and \textsc{rflag} are run independently on all fields. Post-calibration, the bandpass shapes are flatter, which allows the median-filtering approach \textsc{rflag} to be applied more effectively. The thresholds are set lower than the first round as the data is much more well-behaved after calibration. 
\begin{figure}[!h]
    \centering
    \includegraphics[width=\linewidth]{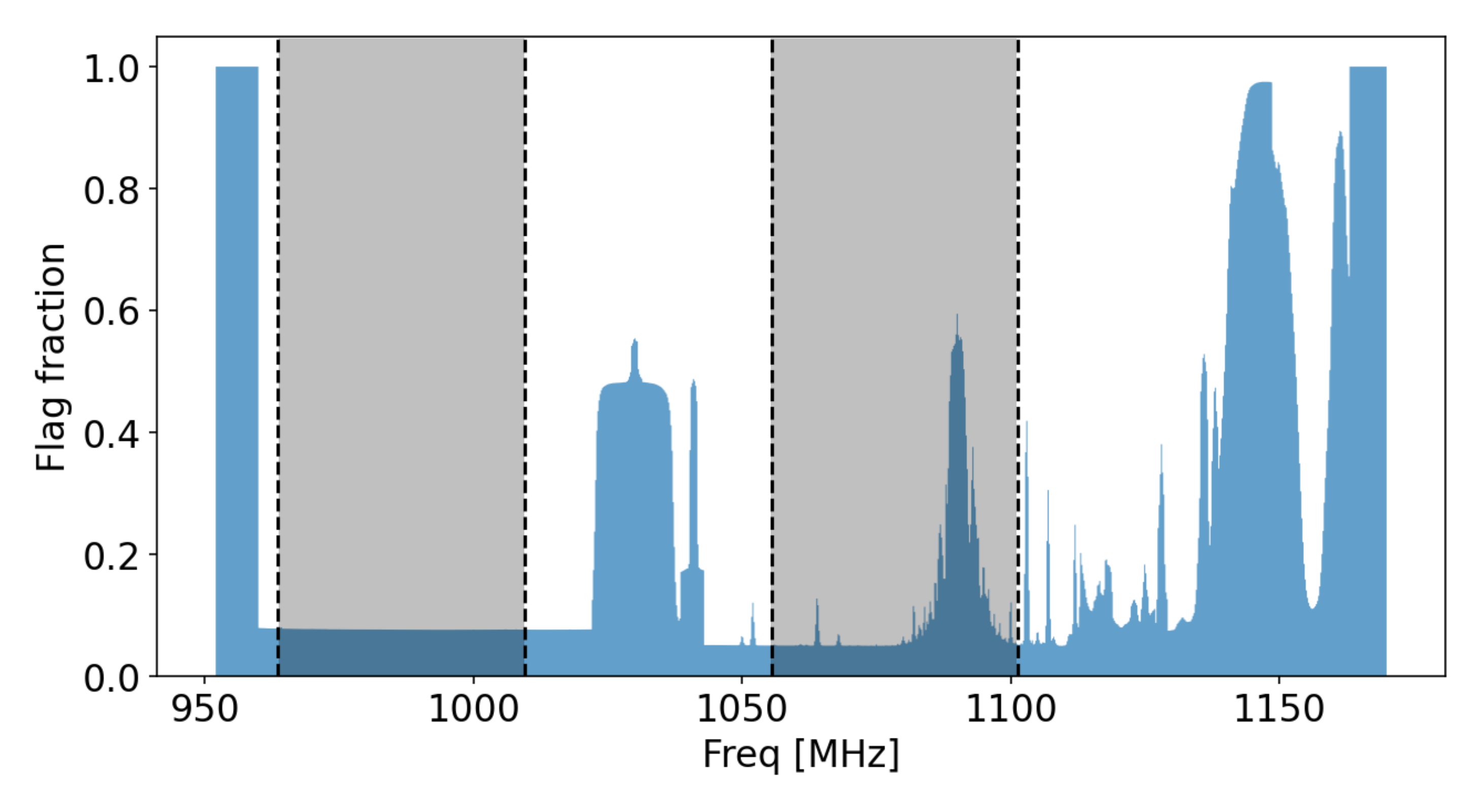}
    \caption{The fraction of flagged data. Only baselines with $-6000\lambda<u,v<6000\lambda$ are selected for calculating the fraction. }
    \label{fig:flag_frac}
\end{figure}

The fraction of data flagged for the baselines within the $uv$ range of our interests is shown in \autoref{fig:flag_frac}. Overall, $\sim 10\%$ of the data is flagged for channels with no severe RFI contamination.
Following this, we perform three rounds of phase-only self-calibration with a solution interval of $60$s. As the target field does not contain any dominant point sources, we do not perform any amplitude self-calibration. 

\begin{figure*}
    \centering
    \includegraphics[width=1.0\linewidth]{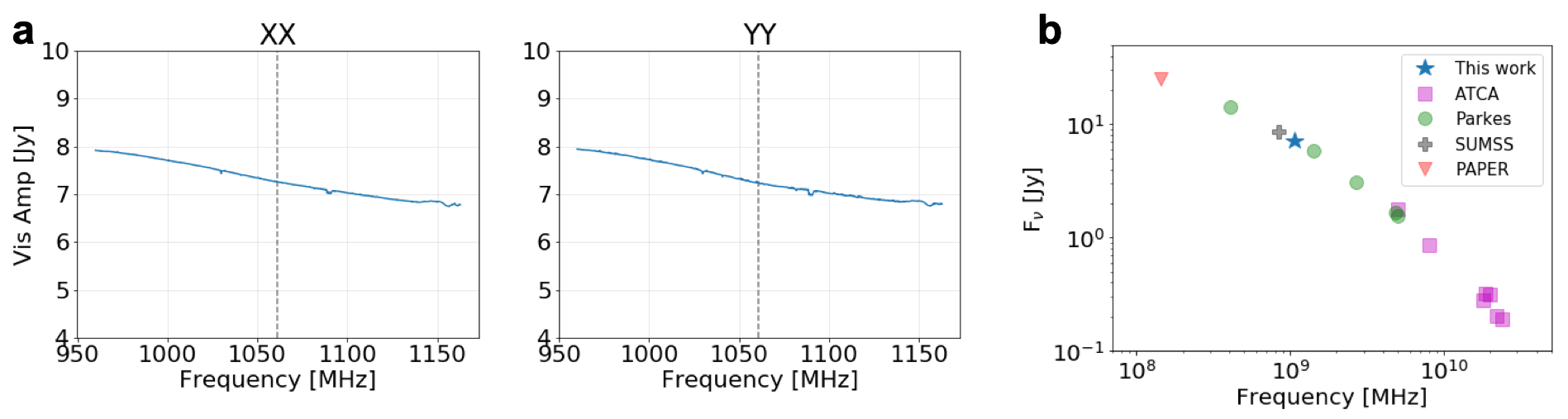}
    \caption{{\bf (a)} Average Visibility amplitude of the secondary calibrator PKS J0252-7104 vs frequency for a single dataset in both XX and YY polarizations. The averaging is done over visibility amplitudes measured by all baselines. {\bf (b)} Comparison with other radio catalogs.}
    \label{fig:flux_0252-712}
\end{figure*}

In \autoref{fig:flux_0252-712}a, we present the visibility amplitude of the secondary calibrator source averaged over all baselines for the dataset id $1530399641$. The Parkes Radio Sources catalogue reports the flux density of $5.9$ Jy at $1410$ MHz \citep{1990PKS...C......0W}. Assuming a spectral index of $-0.7$, the flux density scales to $7.2$ Jy at 1061 MHz (center of the $952{\text -}1170$ MHz sub-band). We measure the flux density as $7.32$ Jy at 1061 MHz (black dashed line in \autoref{fig:flux_0252-712}a and `blue star' marker in \autoref{fig:flux_0252-712}b).

\begin{figure*}
\centering
\includegraphics[width=0.75\textwidth]{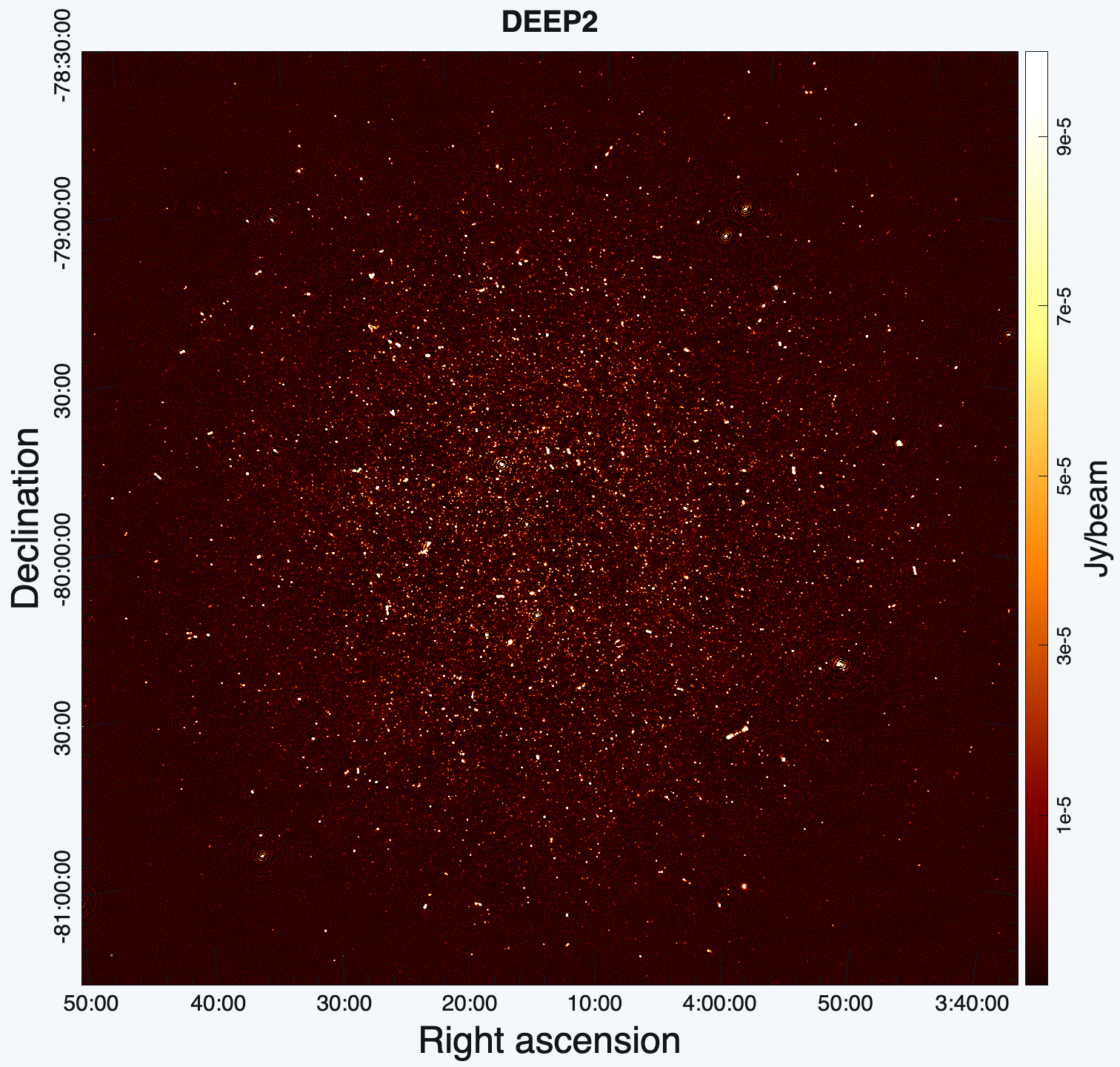} 
 \caption{Stokes I frequency-averaged image of the field at $1077.5$ MHz from 1 ($12.75$ hrs tracking) out of 9 datasets. (Image rendered with CARTA \citep{CARTA})} 
\label{DEEP2}
\end{figure*}

For computing the power spectrum, we adopt a \textit{visibility}-based approach. Therefore, no image cubes are used in our analysis pipeline. Our analysis relies on foreground isolation via the avoidance technique and thus identifying individual foreground sources in the map domain is not required. However, for demonstration purposes, we present a frequency-averaged Stokes~I image of the field at $1077.5$ MHz from a single dataset of $12.75$ hrs tracking using the subband $952-1170$ MHz in \autoref{DEEP2}. The flux scale of the continuum emission of the target field is at the mJy level and shows no obvious artifacts from strong point sources in the far field. Therefore, no corrections for direction-dependent effects had to be applied.

\subsection{Gain stability}
\label{sec:gain}
Calibration errors are a major source of systematics in 21\,cm surveys. It has been established that the calibration bandpass errors need to be $\lesssim 10^{-4}$ for Epoch of Reionization (EoR) experiments to detect the 21\,cm power spectrum \citep{2016MNRAS.461.3135B}. Therefore, it is important to quantify the stability of the gain across frequency and time, as well as the errors of the solutions.

We note that the main source of calibration errors in EoR experiments is usually incomplete sky modelling due to the large field-of-view (FoV) of these instruments, as well as the fact that the short baselines are sensitive to unmodelled extended structures. 
For the MeerKAT telescope, the calibration is done mostly by tracking the calibrator sources, and the effects of field sources on the calibration solution are small due to the $\sim 1\rm $deg beam.
Due to the small FoV and absence of extremely short baselines in MeerKAT, we expect these effects on calibration to be negligible.

\begin{figure}[!ht]
    \centering
    \includegraphics[width=1.0\linewidth]{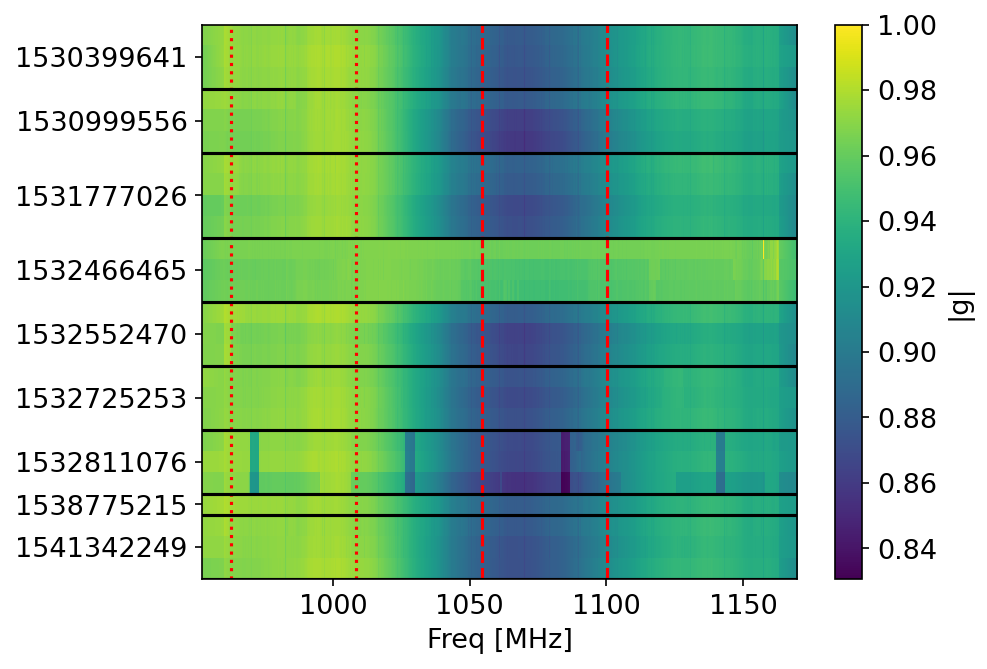}
    \caption{The renormalized bandpass solution averaged across all antennae. The horizontal solid lines separate different observation blocks. The red dotted lines denote the $z=0.44$ bin while the dashed lines denote the $z=0.32$ bin. Each point along the y-axis is a solution interval, and the interval in the middle of a block is denoted with its block id for reference.}
    \label{fig:gain_st}
\end{figure}

In our work, the bandpass calibration is the main source of calibration instability as it is observed less frequently with overall smaller integration times. The solution interval for the bandpass, i.e. the time between two primary calibrator scans, is 3\,hrs. To assess the temporal fluctuations of the gain, we extract the bandpass solutions for both feeds of each antenna. The solutions are then renormalized by the maximum value over the $L$-band frequencies. The solutions are further averaged for both feeds across all antennae for examination. In \autoref{fig:gain_st}, we show the renormalized bandpass solution for each solution interval.

\begin{figure*}[!ht]
    \centering
    \includegraphics[width=1.0\linewidth]{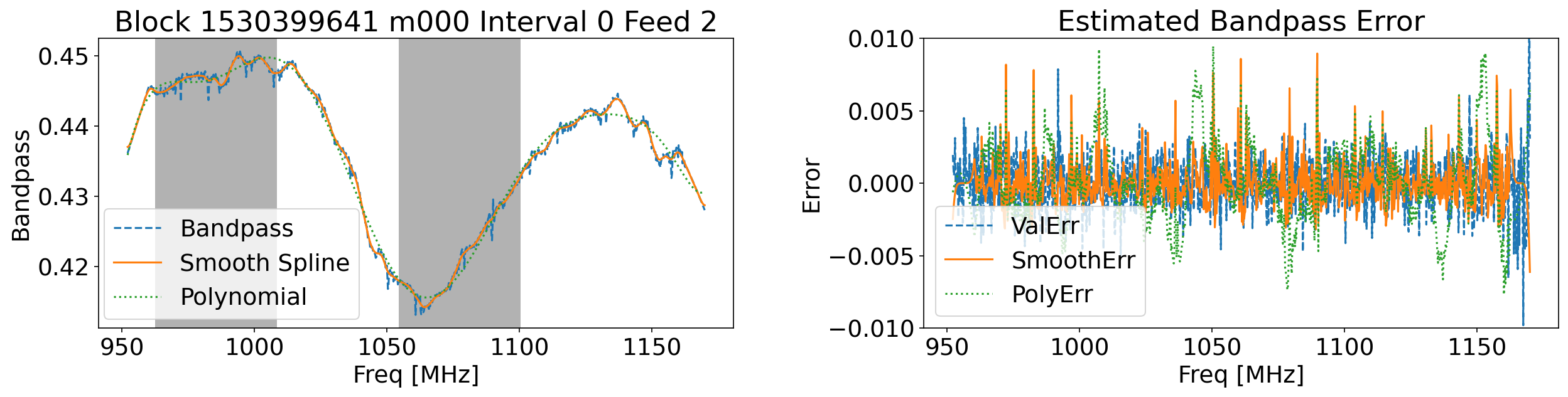}
    \caption{An illustration of the bandpass error estimation. The left panel shows the amplitude of the gain solution for the vertical linearly polarised feed of antenna m000 in the first solution interval of Block 1530399641. The $z\sim 0.32$ and $z\sim 0.44$ sub-bands are marked as the shaded regions. The right panel shows the results of the error estimation, where the y-axis shows the fractional error of the solution, calculated from the \textsc{CASA} solver (``ValErr''), from comparing against the smoothed solution (``SmoothErr''), and from comparing against a tenth-order polynomial fitting (``PolyErr'').}
    \label{fig:gain_err}
\end{figure*}

Overall, we find stable bandpass solutions across the observations, with a smooth frequency structure. In each observation block, the bandpass shape shows negligible variations. However, the solution visibly has dips for some datasets, likely originating from receiver glitches. The effects of the dips are taken into account as we quantify the calibration errors as follows.

Calibration errors are used later to validate the power spectrum pipeline and in \autoref{fig:gain_err}, we present an illustration of how these errors are estimated. For each bandpass solution, the \citealt{CASA} calibration tools return a ``valueErr'' column, which is an estimated error based on the errors of fitting the bandpass solution to the calibrator observed. We have verified that the ``valueErr'' values are dominated by the thermal noise level of the observation. Based on the errors, we generate random Gaussian distribution of bandpass errors, shown as the ``ValErr'' line in the right panel of \autoref{fig:gain_err}. The Gaussian errors fluctuate on the smallest frequency intervals which gives maximum delay scatter. However, they are averaged down when visibility data is gridded across different solution intervals. This does not take into account any possible bias as shown in \autoref{fig:gain_st}. To check for that, we then also fit a smooth spline by applying a Gaussian window with the width of 1\,MHz to the solution. The deviation from the smooth spline to the actual solution is taken to be another estimation of the bandpass error, which we denote as ``SmoothErr''. While ``SmoothErr'' has a smoother frequency structure, if certain frequency channels are consistently biased, the deviations are not averaged down unlike the ``ValErr'' case. For example, it can be seen in Figure \ref{fig:gain_err}, while the ``ValErr'' and ``SmoothErr'' errors have similar amplitude, ``SmoothErr'' contains structures consistent with the visible dips in the bandpass. Note that all the errors shown are renormalised so they are fractional errors. Furthermore, we note that the smooth spline exhibits oscillation structures at $\sim 10\,$MHz scales, corresponding to an expected standing wave caused by reflections in the dish \citep{6410347,2021MNRAS.502.2970A}.

We also produce a conservative estimation of the calibration error where we fit a tenth-order polynomial to the bandpass, see dotted line in \autoref{fig:gain_err}a, and use the deviations as an estimation for the error as marked as `PolyErr' in \autoref{fig:gain_err}b. This pessimistic error estimation includes the oscillation structures in the error.

As we see, overall the gain errors are at the $\sim 0.1\% - 0.5\%$ level. The estimated gain solution errors are then propagated to the rest of the data analysis pipeline, and we find the overall calibration errors to be $\lesssim 0.1\% $ as we discuss later in \autoref{subsec:fgscatter}.

\section{Power Spectrum Pipeline} \label{sec:method}
In this section, we present the formalism to compute the HI power spectrum, which acts as a probe of the underlying dark matter distribution. One of the major challenges in 21-cm cosmology experiments is the overwhelming presence of bright foreground emissions that obscure the weak 21-cm signal. In our study, we use the `delay spectrum' approach \citep{Parsons_2012a, Liu_2014a}, where a Fourier transform is performed along the frequency axis of the gridded visibilities, effectively translating spectral information into delay space and enabling a direct estimation of the 21-cm power spectrum. The delay spectrum technique allows us to separate the foregrounds and the cosmological signal based on their distinct spectral behaviors in Fourier space. The foregrounds such as the Galactic synchrotron emission and radio galaxies are smooth in frequency and contribute power at short delays (small $k_\parallel$). By excising the foreground-dominated modes in the delay spectrum, we can effectively isolate the cosmological signal \citep{Morales_2004, Mcquinn_2006, Datta_2010, Morales_2012, Parsons_2012a, Vedantham_2012, Pober_2013, Parsons_2014, Liu_2014a, Liu_2014b, Paul_2016, 2023MNRAS.518.2971C}. This methodology has proven particularly useful in characterizing the Epoch of Reionization and probing the formation and evolution of the first galaxies \citep{PAPER_2019, HERA_2022, Kolopanis_2023, Gehlot_2018}.

\begin{figure*}
\centering
\includegraphics[width=1.0\textwidth]{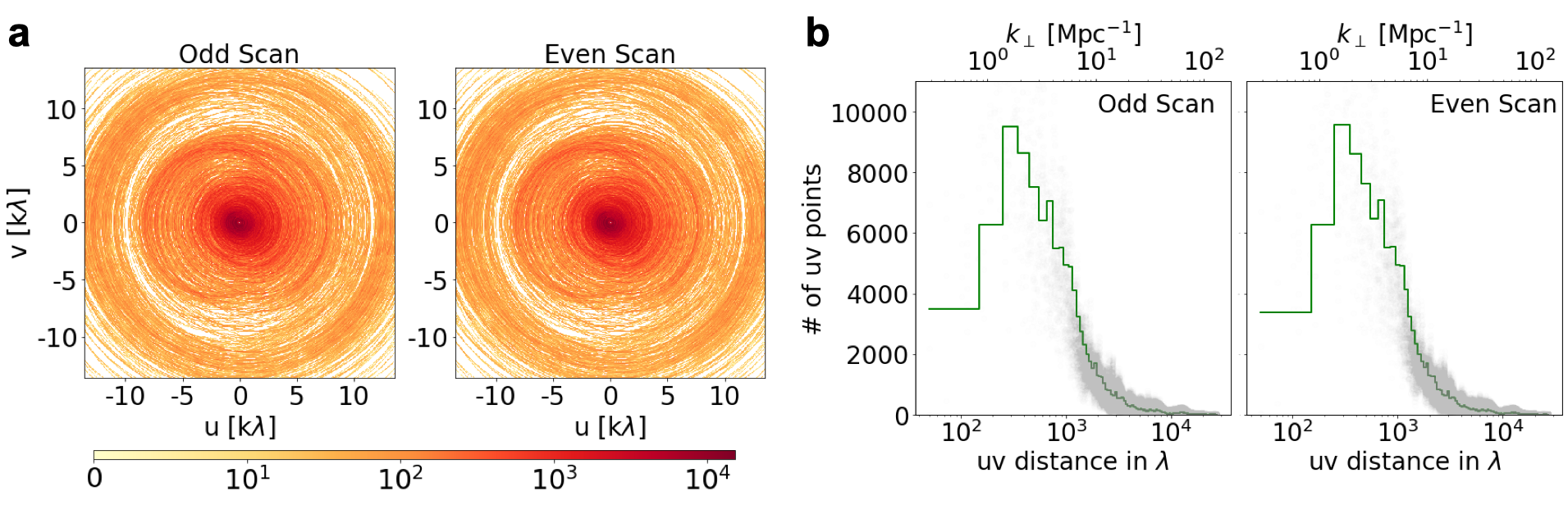}
\caption{{\bf The distribution of $(u,v)$ points on a 2-d plane at $1077.5$ MHz.} The cells on this plane are of size $\Delta u = \Delta v = 60\lambda$. {\bf (a)} The number of $uv$ points on each cell is represented by the colour. {\bf (b)} The number count on each cell is plotted as grey points as a function of $uv$ distance on the lower $x$-axis, whereas the upper $x$-axis denotes the $k_\perp$ values. The green solid line represents the average number of $uv$ points as a function of $uv$ distance (calculated on a bin size of $\Delta uv = 100$).}
\label{uv_dist}
\end{figure*}
\subsection{Visibility Gridding}
We estimate the HI power spectrum using our custom-built pipeline developed in \citet{Paul_2021} which can be referred to for more details. Here, we present a brief summary. The main idea of the delay spectrum approach is to work directly with the interferometer visibilities, instead of producing images and HI cubes. Each baseline with a vector $\bm{b}$ records a visibility $V(\bm{b}, t, \nu)$, which represents the spatial correlation of the electric field for that antenna pair. 
In the limit of small sky areas, $V(\bm{b}, t, \nu)$ measures the 2-d Fourier transform of the sky at each frequency $\nu$ and time $t$. The ``angular" Fourier mode is related to the baseline vector by $\bm{k_\perp} = \frac{2\pi \bm{b}}{\lambda x}$, where $\lambda$ is the observing wavelength, and $x$ is the comoving distance to the redshift corresponding to the observed frequency.
The vector $\bm{b}$ corresponds to a point in the usual $(u,v)$ plane with coordinates $\bm{b}/\lambda$ \citep{Thompson_Moran_book}. In a tracking observation, such as the one used here, the baseline vector will trace a curve in the $(u,v)$ plane as the Earth rotates. Since we are observing the same sky patch, visibilities at a given frequency with the same $(u,v)$ point should have the same value at any given time (except for noise and RFI). We can therefore average together such visibilities, which requires gridding the visibilities in the $(u,v)$ plane.

In order to proceed, we further split the visibility datasets into two smaller frequency bands of $\approx 46$ MHz width centered at $986$ and $1077.5$ MHz respectively ($z \approx 0.44$ and $z \approx 0.32$). The smaller bandwidth segments are chosen to minimize the effects of cosmological evolution ($\Delta z \sim 0.06$). The power spectrum is calculated for these two bands separately. We generate visibility cubes in the baseline ($uv$)-frequency ($\nu$) domain by gridding the $uv$ (or 2-d $\bm{k_\perp}$) plane into discrete cells. The $(u,v)$ coordinates are calculated at the centre frequency of each of the smaller bands. The $(u,v)$ cell size is chosen to avoid decorrelation given the primary beam size ($\sim 52\lambda$ at 1 GHz). The primary beam convolves with the sky signal in Fourier space, mixing different $k_\perp$-modes. In order to minimise the effects of mode-mixing by the primary beam, we choose a large grid size, $\Delta u = \Delta v = 60 \lambda$, corresponding to $\Delta k_\perp \sim 0.3\,{\rm Mpc^{-1}}$. This bin size is kept constant across the frequency since, as mentioned above, the redshift evolution within each of the small bands analyzed is negligible. 

 In our approach, we create two binned datasets from the observations for each frequency band, which can be used to cross-correlate in order to remove the noise bias in the power spectrum. For each observing session, we split the baselines based on the parity of the scan id they belong to. We construct the ``even scan'' and ``odd scan'' visibility gridded cubes from the 9 observing sessions, excluding the flagged baselines.
 We also only consider $uv$ points for which the visibility data has at least $80$ percent unflagged channels (within the $\approx 46$ MHz sub-band) to mitigate potential weak wide-band RFI. Moreover, the flagged channels (if any) are substituted with the nearest neighbour unflagged channel data. This ensures even sampling across channels when we perform the Fast Fourier Transform (FFT) during the delay transformation. The distribution of $(u,v)$ points on the $uv$ grid is shown in \autoref{uv_dist} for both odd and even datasets. Due to a large number of short baselines, the $uv$ pixels are densely populated at smaller $|uv|$ distances, leading to higher sensitivity at smaller $k_\perp$ values.

 Finally, since the sky signal is the same within each $uv$ cell to a good approximation, we can coherently average the visibility data within it. This binning produces two ($uv$)-frequency cubes for each sub-band.

\subsection{3-d Power spectrum}
To calculate the 3-d power spectrum, we start by Fourier transforming the visibilities in the ($uv$)-frequency cubes along the frequency axis (the delay transform). This is done through an FFT along the frequency axis for every $uv$ pixel, producing two $u$-$v$-delay $\tau$ cubes (even and odd). During this delay transformation, the averaged visibility data is multiplied with the Blackman-Harris spectral window function that suppresses the foreground power leakage to higher $k_\parallel$ modes. This results in the delay-space visibility function $\tilde{V}(\bm{k_\perp}, \tau)$. In the limit of small bandwidths, this delay-space can be related to the line of sight Fourier mode, $k_\parallel$, through $k_\parallel = \frac{2\pi\tau\nu_{21}H(z)}{c(1+z)^2}$, where $\nu_{21}$ is the rest frame frequency of the 21-cm line, $H(z)$ is the Hubble parameter at redshift $z$ ($1+z=\nu_{21}/\nu$). This creates the 3-d Fourier cube as a function of $\bm{k}=(\bm{k_\perp}, k_\parallel)$. 

Next, we cross-correlate between even and odd cubes, computing the 3-d $(\bm{k_\perp}, k_\parallel)$ power spectrum  
 \citep{Morales_2004, Mcquinn_2006, Parsons_2012a, Parsons_2014, Paul_2016}:
\begin{linenomath*}
\begin{equation}
P(\bm{k_\perp}, k_\parallel) \equiv  \frac{x^2 y}{\Omega_{\rm ps}B} \left(\frac{\lambda^2}{2k_B}\right)^2 {\rm Re}\{\tilde{V}_1(\bm{b},\tau)\tilde{V}_2^{*}(\bm{b},\tau)\},
\label{PS_eq}
\end{equation}
\end{linenomath*}
with $\lambda$ denoting the wavelength at the band-center, $k_B$ the Boltzmann constant and $y$ the comoving depth corresponding to the bandwidth $B$; the visibility functions $\tilde{V}_1$ and $\tilde{V}_2$ correspond to the odd and even cubes and we are taking the real part (${\rm Re}\{\}$) of the product.
This cross-correlation between different times removes the noise bias in our calculation, and it is also useful to minimize the impact of any time-dependent systematics present in the data.  
 {$\Omega_{\rm ps}$ denotes the power-squared primary beam area, i.e. the solid angle integral of the primary beam squared \citep{2005ApJ...625..575S,Parsons_2014}},
\begin{equation}
    \Omega_{\rm ps} = \int {\rm d}l\,{\rm d}m\, A^2(l,m),
\end{equation}
where $A(l,m)$ is the primary beam attenuation.
The power-squared primary beam area is calculated using the \textsc{eidos} MeerKAT beam model (see \autoref{sec:PrimaryBeam}).

\subsection{Signal 2-d and 1-d Power spectrum}
In the case of statistical isotropy and homogeneity, the power spectrum should be only a function of $k=\sqrt{\lvert \bm{k_\perp} \rvert^2 + k_\parallel^2}$. However, the instrumental noise contribution evolves as a function of $k_\perp$ due to the baseline distribution of the array. The isotropy is also broken due to HI redshift space distortions, which at these small scales tend to reduce the amplitude of fluctuations for large $k_\parallel$ along the line of sight. Nevertheless, the attenuation along the line of sight for the HI signal is much weaker than the attenuation for the foregrounds.

Here, we employ a conservative approach to compute the 1-d power spectrum and only include those $(\bm{k_\perp}, k_\parallel)$ pixels which are considerably above the \textit{horizon limit} of the foreground wedge. For a given angular separation $\theta_0$, from the array pointing, the horizon limit is defined as \citep{2014PhRvD..90b3018L}
\begin{linenomath*}
\begin{equation}
    k_\parallel = \frac{x H(z) \sin \theta_0}{c(1+z)} k_\perp.
    \label{wedge_eq}
\end{equation}
\end{linenomath*}
Due to the small FoV and small sidelobes of the MeerKAT telescope, the horizon limit set by the primary beam is close to $k_\parallel \sim 0.02 k_\perp$. However, this does not exclude modes that are susceptible to foreground leakage from the sidelobes and instrument systematics. Assuming the most extreme case that leakage of bright foreground sources from the side lobes extends to the entire sky, i.e. $\sin \theta_0 = 1$, the horizon limit becomes $k_\parallel \sim 0.26 k_\perp$. Therefore, we use a much stricter selection criterion $k_\parallel = 0.3 k_\perp$ to ensure the robustness of our results.

From the 3-d cylindrical power spectrum $P(\bm{k_\perp}, k_\parallel)$, we use inverse noise variance weighting ($1/\sigma^{2}(\bm{k_\perp}, k_\parallel)$) to calculate the optimal 2-d power spectrum $P(k_\perp, k_\parallel)$ with $k_\perp=|\bm{k_\perp}|$ and 1-d power spectrum $P(k)$ with $k=|(\bm{k_\perp}, k_\parallel)|$. The weighted average for both cases is:
\begin{equation}
    \hat{P}^{i} = \big(\sum_j P^{ j} w_{ j} \big)/\big(\sum_j w_{ j}\big),
\label{eq:avg}
\end{equation}
where $j$ loops over all the $k$-points that fall into the $i^{\rm th}$ $k$-bin and $w_{ j} = 1/\sigma^2_j$ is the inverse covariance weight calculated from thermal noise simulations.
The errors are then calculated from the sampling variance of the 3-D powers in each bin:
\begin{equation}
    (\Delta P^{\rm i})^2 = \big(\sum_j (P^{\rm j}-\hat{P}^{\rm i})^2 w_{\rm j}^2 \big) \Big/ \big(\sum_j w_{\rm j}\big)^2.
\label{eq:err}
\end{equation}

\begin{figure*}
\centering
\includegraphics[trim={0 70 0 40},clip,width=1.0\textwidth]{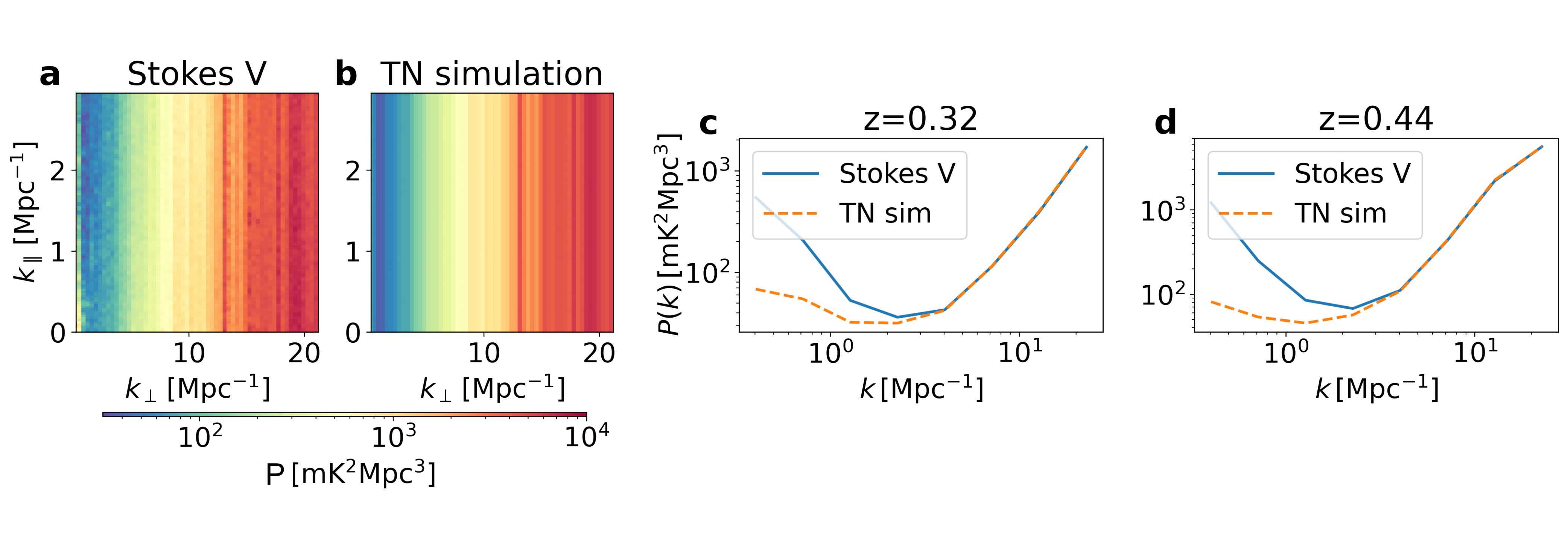}
\caption{{\bf (a)} The cylindrical auto power spectrum of the Stokes V mode calculated from the $z=0.44$ data (odd-scan cube). {\bf (b)} Simulated thermal noise auto power spectrum averaged across 10000 realizations. {\bf (c)} The 1-d auto power spectrum of the Stokes V mode compared against the thermal noise simulation for $z=0.32$. {\bf (d)} The same for $z=0.44$. }
\label{2d_ps_TN}
\end{figure*}

\subsection{The Noise Power Spectrum}
\label{sec:Noise_sims}
We estimate the instrument noise contribution to the power spectrum variance by generating cylindrical thermal noise power spectrum following the description in \cite{Paul_2021}. For the cylindrical noise power spectrum, each calibrated visibility component is replaced with a simulated thermal noise complex visibility generated from a zero mean random process with standard deviation \citep{Taylor_1999}
\begin{linenomath*}
\begin{equation}
\sigma_{\rm{N}} = \frac{2k_{\rm{B}} T_{\rm{sys}}}{A_e\sqrt{\delta\nu \delta t}},
\label{sigma_N}
\end{equation}
\end{linenomath*}
where $T_{\rm{sys}}$ and $A_e$ are the system temperature and effective area of each antenna respectively, $k_{\rm{B}}$ is the Boltzmann constant, $\delta\nu = 208.984$ kHz is the channel width and $\delta t = 8$s is the time resolution. The thermal noise amplitude depends on the natural sensitivity of the instrument $A_e/T_{\rm sys}$. We estimate the natural sensitivity at each frequency sub-band by calculating the variance of the Stokes V visibility data. The Stokes V mode is a good estimator of the thermal noise as the extragalactic foreground sources have minimal circular polarization. For MeerKAT observations, the Stokes V data is consistent with the thermal noise at long baselines \citep{Paul_2021}. We use the same gridding routine as in \autoref{sec:method} to grid the Stokes V data without filling the flagged channels with the nearest neighbour, and use the gridded visibility to calculate the Stokes V power spectrum. 

Based on the number of baselines in each $u$-$v$ grid, we can calculate the thermal noise amplitude $\sigma_N$
\begin{equation}
    \sigma_{N_i} = \sqrt{N_i}\ {\rm std}\left({\rm Re}\{V_i\}\right),
    \label{Esigma_N}
\end{equation}
where $N_i$ and $V_i$ are the (non-zero) number of baselines and averaged Stokes~V visibility respectively within the $i$th $u$-$v$ grid. The standard deviation is computed across the channels assuming the system temperature is constant across the (small) band. This rms is then averaged over the $u$-$v$ grid cells. However, only $u$-$v$ points corresponding to modes $k_\perp > 10$ Mpc$^{-1}$ are used. This is to guarantee that the Stokes V signal is noise dominated.
Finally, the natural sensitivity $A_e/T_{\rm sys}$ is calculated by comparing \autoref{sigma_N} with \autoref{Esigma_N}. We find $A_e/T_{\rm sys} = 6.65\,{\rm m^2/K}$ for $z=0.32$ and $6.48\,{\rm m^2/K}$ for $z=0.44$, consistent with the anticipated performance of $A_e/T_{\rm sys} = 6.22\,{\rm m^2/K}$\footnote{MeerKAT array specification document as on 2016-10-10, available in: \url{https://www.sarao.ac.za/science/meerkat/ about-meerkat/}}. We then generate $10,000$ realizations of the noise 3-d power spectrum using this measured $A_e/T_{\rm sys}$. The variance of the power spectrum due to noise is then calculated as the variance over these realisations at each $k-$pixel, $\sigma^2(\bm{k_\perp}, k_\parallel)$. For a given $\bm{k_\perp}$, $\sigma$ is approximately constant across $k_\parallel$, only changing across $\bm{k_\perp}$ depending on the baseline density.

\autoref{2d_ps_TN} depicts the power spectra of the Stokes V data from the odd-scan compared to the one resulting from averaging over our thermal noise (TN) simulations employing the same $uv$ distribution of the odd-scan data (and the same horizon cut). We show a comparison of the cylindrical power spectrum on the left for $z=0.44$, and the spherical power spectra for both redshift bins on the right. The Stokes V power spectrum agrees well with our thermal noise simulation, especially at long baselines (high $k_\perp$). At shorter baselines, i.e. short $k_\perp$ shown as vertical lines in the left of the plot, the increasing level of systematics and polarization signal breaks the similarity.

\begin{figure*}
    \centering
    \includegraphics[trim={0 0 0 0},clip,width=1.0\linewidth]{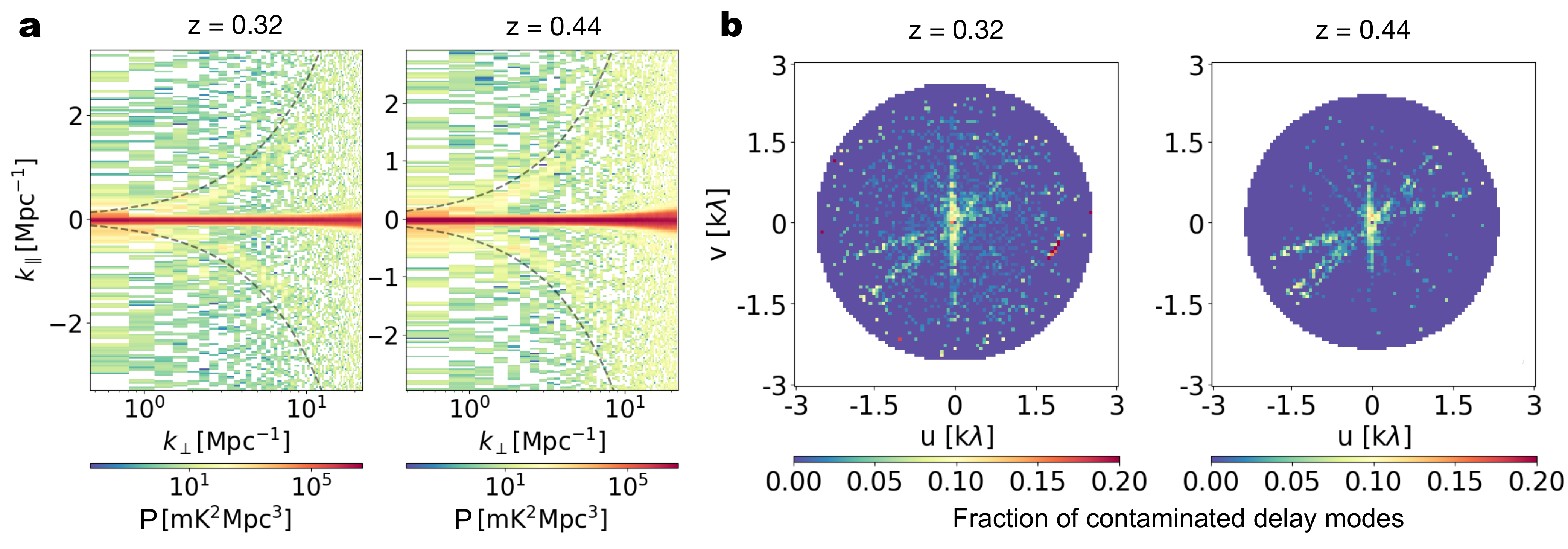}
    \caption{{\bf Indication of Low-Level broadband RFI: }{\bf (a)} The Stokes I 2-d power spectrum derived from the analysis of $\mathbf{96}$ hours of MeerKAT interferometer data at $\mathbf{z=0.32}$ and $\mathbf{z=0.44}$, cross-correlating the even and odd scan visibility cubes. Negative values are omitted. The dashed line marks the foreground-avoidance boundary $k_\parallel = 0.3\,k_\perp$, corresponding to the expected horizon (wedge) extent plus an additional buffer; only modes above this line are used in the 1d power-spectrum analysis. Stripes of excess power are visible near this boundary in both redshift bins. {\bf (b)} After $uv$ gridding and performing FFT along frequency, the fraction of delay modes which exhibit power spectrum amplitude exceeding $5\sigma$ of the thermal noise is shown in color for both redshift bins. This fraction is significantly higher near the $u=0$ line relative to other regions of the $uv$ plane. These affected modes contribute to the excess power observed in the 2-d power spectrum.}
    \label{fig:2d_PS_1}
\end{figure*}

\section{Results} \label{sec:results}

\subsection{The 2-d power spectrum and presence of low-level RFI}
\label{sec:RFI}
In \autoref{fig:2d_PS_1}a, we show the 2-d power spectrum (Stokes I) as a function of $(k_\perp, k_\parallel)$ at $z=0.32$ and $z=0.44$ by cross-correlating the odd and even scan cubes. The dashed line defines the horizon limit of the wedge, $k_\parallel = 0.3 k_\perp$, above which we do not expect any foreground contamination (e.g. \citealt{Morales_2012}). In our case, the superb stability of MeerKAT and its primary beam size move the foreground contaminants to a much smaller region. Nevertheless, we take a conservative approach and restrict our analysis to values above the dashed line. This foreground avoidance technique has the advantage of being robust to signal loss. Apart from the foreground-dominated modes at low $k_\parallel$, the rest of the cylindrical $k$-space is expected to be noise-dominated. While most regions beyond the foreground wedge in \autoref{fig:2d_PS_1}a are noise-dominated, we detect stripes of excess power near the $k_\parallel=0.3k_\perp$ line. To investigate their origin, we examine the voxels in the gridded $uv-\tau$ domain where the power spectrum amplitude exceeds the expected thermal noise. Specifically, we examine the power spectrum amplitudes for a given delay mode in the gridded visibility data from the odd and even cubes, identifying contaminated modes if the amplitude exceeds $5\sigma$ of the expected thermal noise in either cube. In \autoref{fig:2d_PS_1}b, we show the fraction of all delay modes, beyond the horizon boundary (black dashed lines in \autoref{fig:2d_PS_1}a), that are classified as contaminated across the $uv$ plane. This fraction is significantly higher near the $u=0$ line, indicating the likely presence of weak broadband RFI that went undetected during the initial flagging process. This has been observed before in MeerKAT data\footnote{\url{https://github.com/caracal-pipeline/caracal/discussions/1398}} and is related to the reduced smearing from fringe rotation towards $u=0$. As an additional diagnostic, we explored whether the excess power could be traced to a localized broadband emitter near the horizon (e.g., a stationary RFI source or transient lightning). We therefore formed wide-field images extending to the horizon using the calibrated visibilities. However, we did not identify any localized horizon source (persistent or transient) that correlates with the $u=0$ systematics. While this does not rule out extremely faint, intermittent, or spatially distributed interference that may remain below the detectability of snapshot imaging (particularly given the strong primary-beam attenuation away from target), it disfavors a single dominant localized emitter as the principal driver of the observed delay-domain excess.

In the following sections, we outline two strategies for detecting and mitigating these systematics.

\begin{figure}
    \centering
    \includegraphics[trim={0 20 0 15},clip,width=1.0\linewidth]{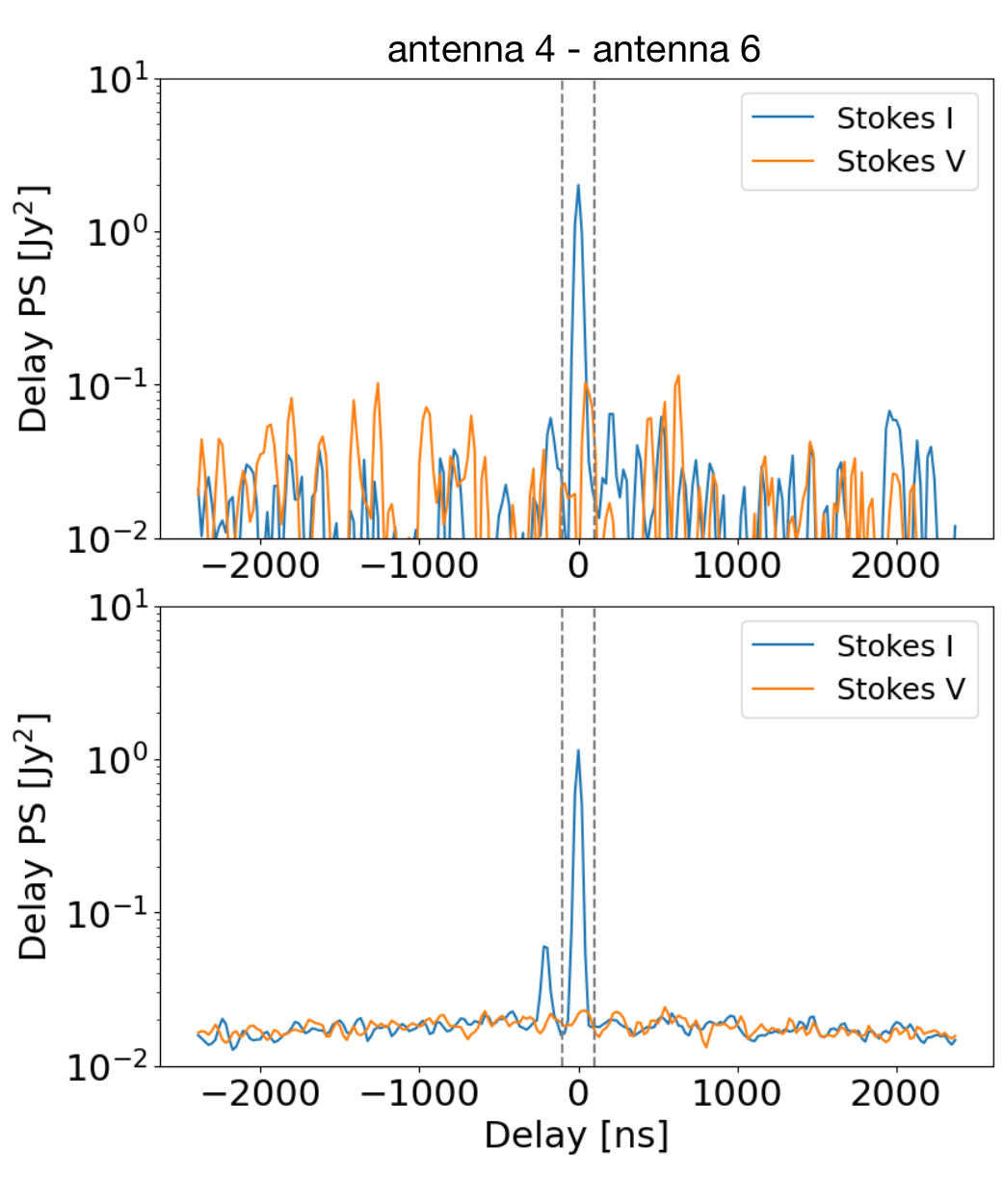}
    \caption{Delay Power spectra for an antenna pair affected by broadband RFI during one particular scan ($z=0.32$ band). {\bf Top panel}: for one 8s time integration, {\bf bottom panel:} averaged over all time samples within a 15-minute scan. The region bounded by the vertical dashed lines ($\pm 100$ ns) denotes foreground-affected modes. The Stokes V power is added to give an estimate of the expected thermal noise. The contamination cannot be identified in the top plot as noise dominates the higher delay modes. Upon averaging all time samples within a scan, the thermal noise drops substantially and the contamination can be identified on the bottom plot (small peak at negative delay on the left of the foreground-dominated region).}
    \label{fig:bl_flagging_1}
\end{figure}

\subsection{Flagging at baseline level (BLF)}
\label{sec:bl_flagging}
The contamination due to the leftover RFI is weak and difficult to distinguish in the per-baseline visibility data. The top panel of \autoref{fig:bl_flagging_1} shows the delay power spectrum for a single 8-second time integration from an antenna pair affected by these systematics. The power outside the foreground wedge primarily reflects noise. However, upon averaging it over a 15-minute scan ($\sim 112$, 8-second samples), we observe peaks at non-zero delays (bottom panel of \autoref{fig:bl_flagging_1}), indicating contamination from broadband RFI. We note that simply averaging the visibilities in time would not work because it would also smear out the RFI. Furthermore, exploratory analyses of these affected baselines in fringe-rate/delay space reveal that this anomalous power is concentrated close to zero fringe-rate. This specific temporal and delay-space distribution supports the interpretation of these systematics as faint, terrestrial RFI or instrumental mutual coupling, rather than celestial emission.

In order to proceed, we analyse the delay power spectrum for every antenna pair, incoherently averaged over each of the 15-minute scans. If any peaks are detected outside the foreground wedge, the corresponding antenna pair is classified as bad and flagged for that specific scan. Peaks are defined with a 5-sigma threshold above the expected thermal noise power. We found that the 15-minute averaging is a good balance since the contamination per baseline is not constant across time. This technique operates independently of the power spectrum estimation pipeline and is applied after calibration but before calculating the power spectrum. This baseline flagging approach (hereafter referred to as BLF) can be likened to other flagging routines commonly used in radio interferometric data processing. 

\begin{figure*}
    \centering
    \includegraphics[trim={0 0 0 0},clip,width=1.0\linewidth]{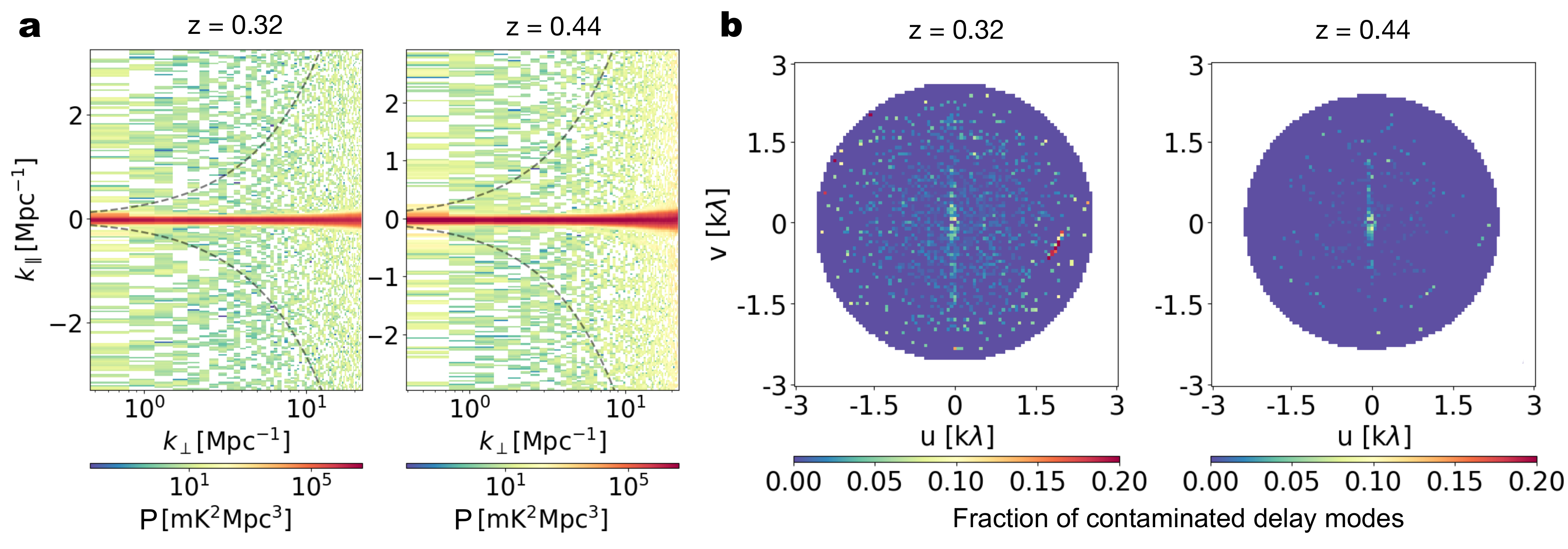}
    \caption{{\bf Filtering broadband RFI systematics:} With the baseline level flagging (BLF), {\bf (a)} the Stokes I 2-d power spectrum for both redshift bins and {\bf (b)} fraction of delay modes above thermal noise on each grid cell. A comparison with \autoref{fig:2d_PS_1} shows a considerable improvement in filtering out the systematics. Although, some residual contaminations along the $u=0$ are still present.}
    \label{fig:2d_PS_2}
\end{figure*}

\begin{figure}
    \centering
    \includegraphics[trim={0 70 0 30},clip,width=1.0\linewidth]{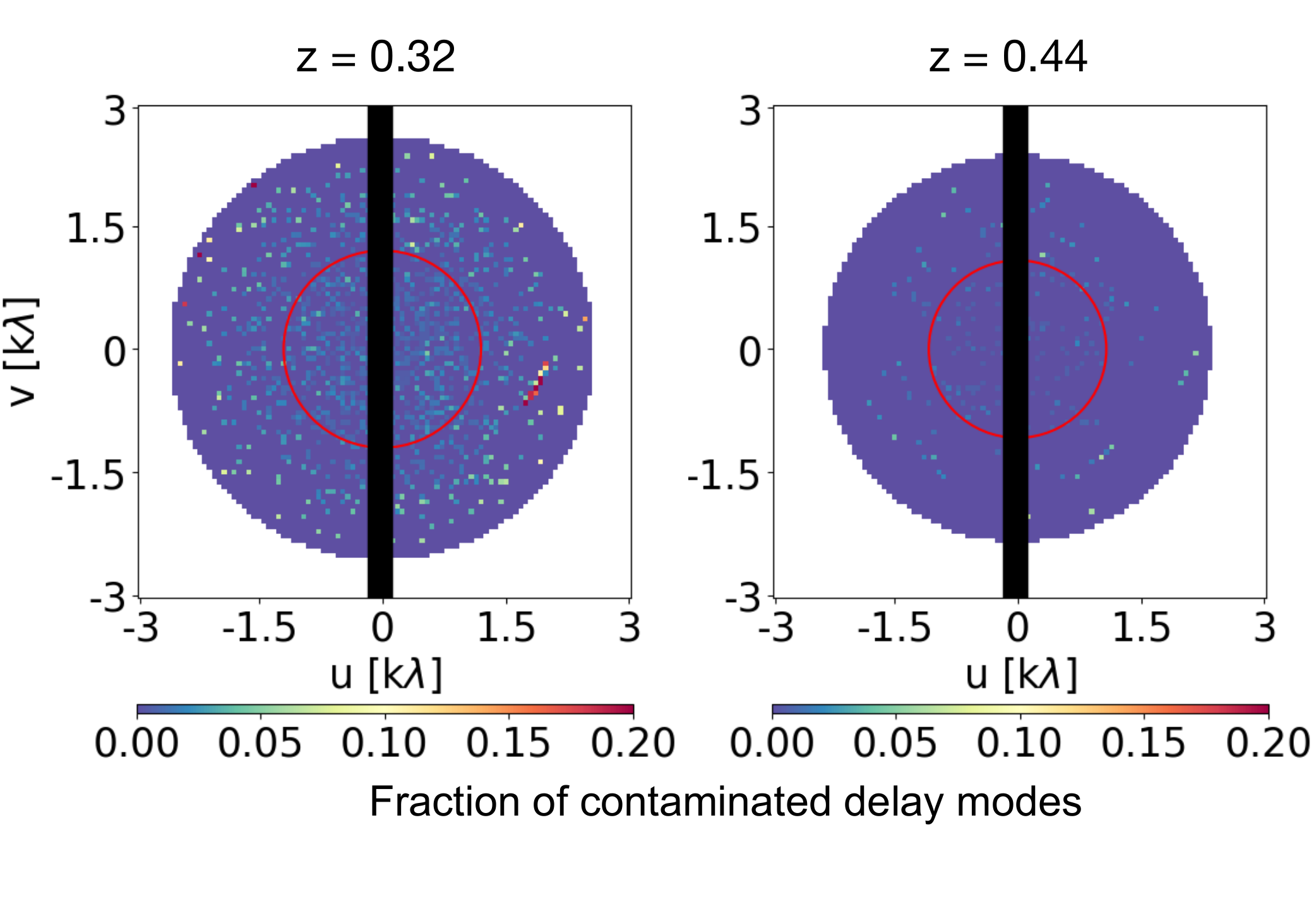}
    \caption{Same plot as \autoref{fig:2d_PS_2}b with a black mask on the $uv$ plane to avoid residual systematics. The $uv$ modes outside this mask and within the red circle are used to calculate the 1-d power spectrum using BLF.}
    \label{fig:uv_mask}
\end{figure}

\begin{figure*}
    \centering
    \includegraphics[trim={0 20 0 20},clip,width=0.95\linewidth]{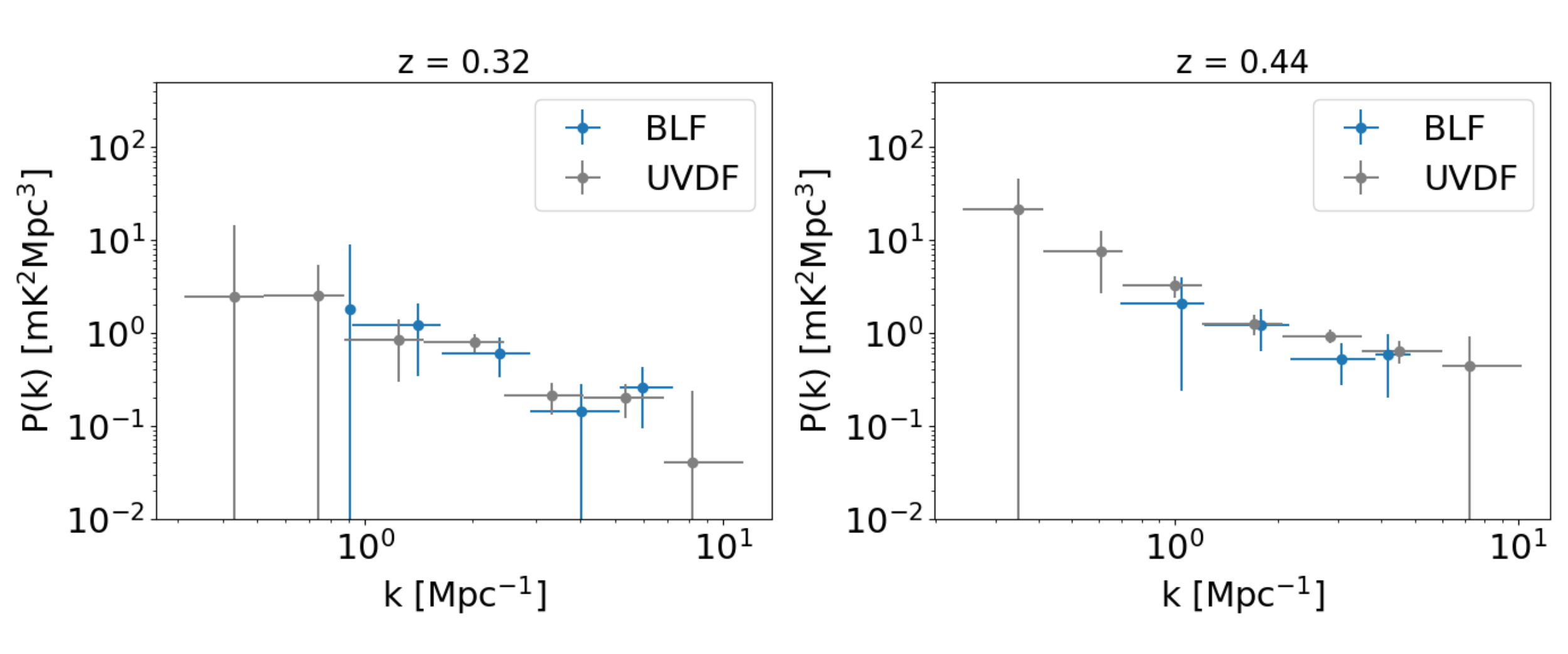}
    \caption{{\bf 1-d Power Spectrum} from the analysis of $\mathbf{96}$ hours of MeerKAT interferometer data at $\mathbf{z=0.32}$ and $\mathbf{z=0.44}$. The mitigation of residual RFI (as discussed in \autoref{sec:results}) is achieved through the implementation of two distinct approaches: BLF (\autoref{sec:bl_flagging}) and UVDF (\autoref{sec:PS_flagging}).}
    \label{fig:1d_PS}
\end{figure*}

In \autoref{fig:2d_PS_2}, we present the results of applying the flagging technique to the data. It effectively filters out most contaminated baselines, as evidenced by the substantial improvement in the 2-d power spectrum in \autoref{fig:2d_PS_2}a compared to \autoref{fig:2d_PS_1}a. However, some residual systematics persist, particularly along the $u=0$ line and in the top-right region of the $uv$ plane, as seen in \autoref{fig:2d_PS_2}b. To calculate the final 1-d power spectrum and increase the statistical significance, we apply a mask to exclude the $uv$ modes near $u=0$ as depicted by the black patch in \autoref{fig:uv_mask} and further average the power spectrum into logarithmic bins in $k$ with inverse covariance weighting described in \autoref{eq:avg} and \autoref{eq:err}. Additionally, we confine the power spectrum estimation within the boundaries of the red circle to mitigate the impact of contaminated $uv$ pixels located in the top-right area. The points outside the circle have high noise levels (low number of baselines) and would contribute little to the final HI signal in any case. The resulting 1-d power spectrum is shown in \autoref{fig:1d_PS}.

\begin{figure*}
    \centering
    \includegraphics[trim={0 30 0 70},clip,width=\linewidth]{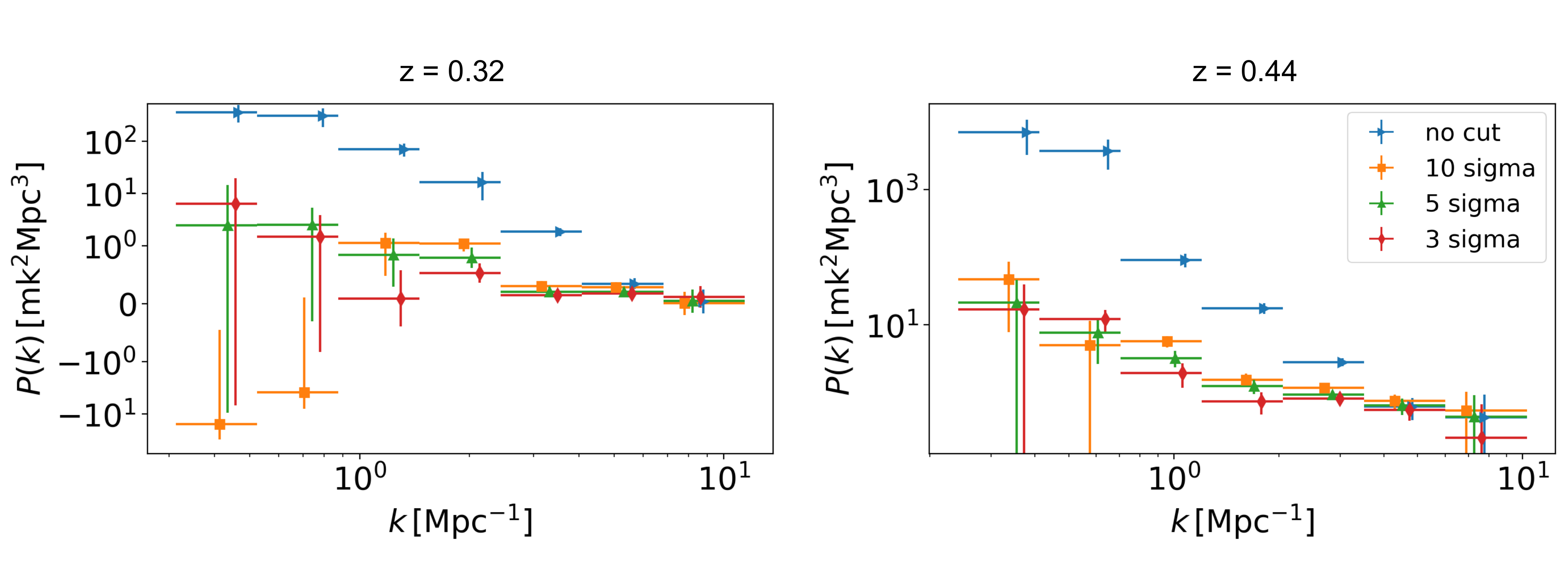}
    \caption{The measured 1-d power spectrum with no thresholding, $10\sigma$, and $5\sigma$ and $3\sigma$ cut (the flags are applied at the auto-power level for each delay transformed visibility cube). The centres of the $k$-bins are misplaced for better visualization. The $5\sigma$ cut is used in our final results with UVDF and the first two $k$-bins have very low signal-to-noise.}
    \label{fig:pssigmacompare}
\end{figure*}

\subsection{Flagging contaminated modes on gridded $uv-\tau$ space (UVDF)}
\label{sec:PS_flagging}
Since the easiest way to see this contamination is through outliers in the gridded $uv-\tau$ space, one might reason that flagging directly these outliers would be an obvious approach. This method might raise some concerns that we discuss in \autoref{sec:crit_uvdf}. Here we present the technique, hereafter referred to as UVDF, an acronym for $uv$-delay flagging, and the corresponding results.

 In this approach, we use thermal noise simulations to identify and flag extreme outliers in the 3-d delay power in $uv-\tau$ space. Specifically, we flag $\bm{k}$ pixels for which $|P(\bm{k_\perp}, k_\parallel)-P_{\rm TN}(\bm{k_\perp}, k_\parallel)| > 5\sigma(\bm{k_\perp}, k_\parallel)$, thereby excluding $k$-modes affected by systematics. Here $P(\bm{k_\perp}, k_\parallel)$ and $P_{\rm TN}(\bm{k_\perp}, k_\parallel)$ represent the auto-power spectrum for Stokes I and thermal noise respectively.  Note that the thermal noise simulations (see \autoref{sec:Noise_sims}) are performed for the gridded odd and even visibility data sets separately, as different time blocks and Stokes parameters at different frequency sub-bands have varying $u$-$v$ sampling. We apply the aforementioned selection criteria to the 3-d auto-power of the Stokes I visibility on each sub-band individually. If a 3-d $\mathbf{k}$-pixel is flagged in either the even or odd scans, it is excluded from the power spectrum estimation in the cross-power. We find that approximately $1\%$ of the $\mathbf{k}$-pixels are flagged by UVDF within the $k_\parallel > 0.3 k_\perp$ window, most of which fall within the stripe structure shown in \autoref{fig:2d_PS_1}a. Our simulations confirm that this selection criterion leads to negligible signal loss (\autoref{signal_recovery}). The resulting 1-d power spectrum is shown in \autoref{fig:1d_PS}. Although UVDF occurs in the later stages of power spectrum estimation, it only flags the contaminated 3-d $\mathbf{k}$-pixels, allowing for the retention of more data and thus improving the sensitivity of the measurements. In comparison, the BLF approach is more aggressive, flagging all data associated with the problematic antenna pair, not just the contaminated delay modes.

\subsection{Assessment of UVDF flagging}
\label{sec:crit_uvdf}
The UVDF flagging is performed on the 3d power spectrum cells and consequently, it could be argued that the flagging is performed ``too close''  to the final measurement product. For instance, assuming a measurement over a quantity $A_i$ and, similarly to the UVDF flagging, outlier values are removed using the expected distribution for $A_i$, there could be the danger to artificially forcing $A_i$ to follow an expected distribution which does not reflect the data's underlying statistics. On the other hand, it would likely be acceptable if only a few extreme outliers, $A_i$, were removed. 
The main concerns regarding the flagging process are signal loss and residual RFI contamination, which are discussed in the following. 

We remind the reader that the $5\sigma$ threshold is calculated using the noise simulations with $A_{\rm e}/T_{\rm sys}$ fitted to the Stokes V values from long baselines. Moreover, since the noise dominates the HI signal at every $\mathbf{k}$ pixel, its distribution should be an accurate representation of the expected values when there is no contamination. Using more aggressive thresholds below $5\sigma$ could lead to signal loss. This could in principle be compensated by a transfer function, leading to a trade-off with increased error bars. However, a transfer function would also increase the possibility of biases and therefore $5\sigma$ is an appropriate choice for this data. Using simulations, as presented in \autoref{signal_recovery}, we verified that the impact of the UVDF flagging on signal loss is negligible. 

We also investigate if higher $\sigma$ levels would be sufficient to flag the data. \autoref{fig:pssigmacompare} illustrates the final power spectrum without any flagging of $\mathbf{k}$ pixels, as well as with $10\sigma$, $5\sigma$ and $3\sigma$ threshold levels. Although there are still some visible contaminants on the 2-d power spectrum, the values of the 1-d power spectrum with a $10\sigma$ cut are relatively consistent with the $5\sigma$, except for the first two bins, which we know suffer from more contamination and have very low signal-to-noise. {Overall, comparing the different choices with the fiducial 5-sigma UVDF measurements, we find a reduced chi-square of [28, 0.45, 0.33] for no-cut, 10-sigma and 3-sigma, respectively.} The consistency with the BLF approach gives us confidence that a non-biased detection is being made.

There is still the possibility that some contaminant could still be responsible for the measured power. Clearly RFI is pervasive on Earth and will be present at some level in the data, e.g. \citet{2023ApJ...957...78W}. However, some of this RFI should integrate down with time, and the cross-correlation approach should further reduce the impact of the RFI. The null tests presented in \ref{validation-null} are further evidence that the final power spectra are dominated by the cosmological signal.

\subsection{Statistical significance of 1-d power spectrum}
\label{sec:1d_ps_comparison}
The estimated 1-d power spectrum values corresponding to \autoref{fig:1d_PS} are provided in \autoref{ps_table_BLF} and \autoref{ps_table}, using the BLF and UVDF methods, respectively. For both redshifts, the 1-d power spectrum shows a clear detection. With the BLF method, the statistical significance is $3.2\sigma$ at $z=0.32$ and $3.5\sigma$ at $z=0.44$, while for the UVDF method, these values are $6\sigma$ and $9.3\sigma$, respectively. The first two $k$-bins in the UVDF results are more susceptible to systematics, as they are automatically excluded by the mask selection in BLF (\autoref{fig:uv_mask}). Excluding these two bins, the statistical significances with UVDF are $5.9\sigma$ and $9.18\sigma$ for the two redshift bins, respectively.
\begin{table*}[!t]
\centering
\begin{tabular}{| c | c c c | c | c c c |}
 \hline
 \multicolumn{4}{|c|}{$z=0.32$} & \multicolumn{4}{c|}{$z=0.44$} \\
 \hline
 $k$[Mpc$^{-1}$] & $P(k)$[mK$^2$Mpc$^3$] & $\sigma_{P}$[mK$^2$Mpc$^3$] & $P(k)/\sigma_{P}$ & $k$[Mpc$^{-1}$] & $P(k)$[mK$^2$Mpc$^3$] & $\sigma_{P}$[mK$^2$Mpc$^3$] & $P(k)/\sigma_{P}$\\
 \hline
 0.91 & 1.79 & 7.07 & 0.25 & 1.05 & 2.07 & 1.83 & 1.13 \\
 1.40 & 1.20 & 0.86 & 1.39 & 1.78 & 1.21 & 0.57 & 2.14 \\
 2.38 & 0.60 & 0.28 & 2.19 & 3.06 & 0.52 & 0.25 & 2.09\\
 4.03 & 0.14 & 0.13 & 1.07 & 4.19 & 0.59 & 0.39 & 1.53\\
 5.95 & 0.26 & 0.16 & 1.58 & - & - & - & -\\
 \hline
\end{tabular}
\caption{Summary of the HI power spectrum constraints at $z=0.32$ and $z=0.44$ using BLF method described in \autoref{sec:bl_flagging}. From left to right, the columns show the centre of each $k$-bin, the value of the measured HI power spectrum in the units of mK$^2$Mpc$^3$, the measurement uncertainties in the units of mK$^2$Mpc$^3$, and the significance of the detection in each $k$-bin.}
\label{ps_table_BLF}
\end{table*}

\begin{table*}[!t]
\centering
\begin{tabular}{| c | c c c | c | c c c |} 
 \hline
 \multicolumn{4}{|c|}{$z=0.32$} & \multicolumn{4}{c|}{$z=0.44$} \\
 \hline
 $k$[Mpc$^{-1}$] & $P(k)$[mK$^2$Mpc$^3$] & $\sigma_{P}$[mK$^2$Mpc$^3$] & $P(k)/\sigma_{P}$ & $k$[Mpc$^{-1}$] & $P(k)$[mK$^2$Mpc$^3$] & $\sigma_{P}$[mK$^2$Mpc$^3$] & $P(k)/\sigma_{P}$\\
 \hline
 0.43 & 2.45 & 11.98 & 0.20 & 0.35 & 21.29 & 24.46 & 0.87 \\
 0.74 & 2.52 & 2.83 & 0.89 & 0.61 & 7.60 & 4.95 & 1.53 \\
 1.24 & 0.84 & 0.54 & 1.55 & 1.00 & 3.23 & 0.87 & 3.69\\
 2.03 & 0.80 & 0.18 & 4.54 & 1.70 & 1.25 & 0.30 & 4.18\\
 3.32 & 0.21 & 0.08 & 2.44 & 2.84 & 0.92 & 0.15 & 6.28\\
 5.32 & 0.20 & 0.08 & 2.41 & 4.52 & 0.64 & 0.18 & 3.59\\
 8.22 & 0.04 & 0.20 & 0.22 & 7.25 & 0.44 & 0.47 & 0.93\\
 \hline
\end{tabular}
\caption{{Same as \autoref{ps_table_BLF} with UVDF approach (\autoref{sec:PS_flagging})}}
\label{ps_table}
\end{table*}

\begin{figure*}
\centering
\includegraphics[trim={0 30 0 30},clip,width=1.0\textwidth]{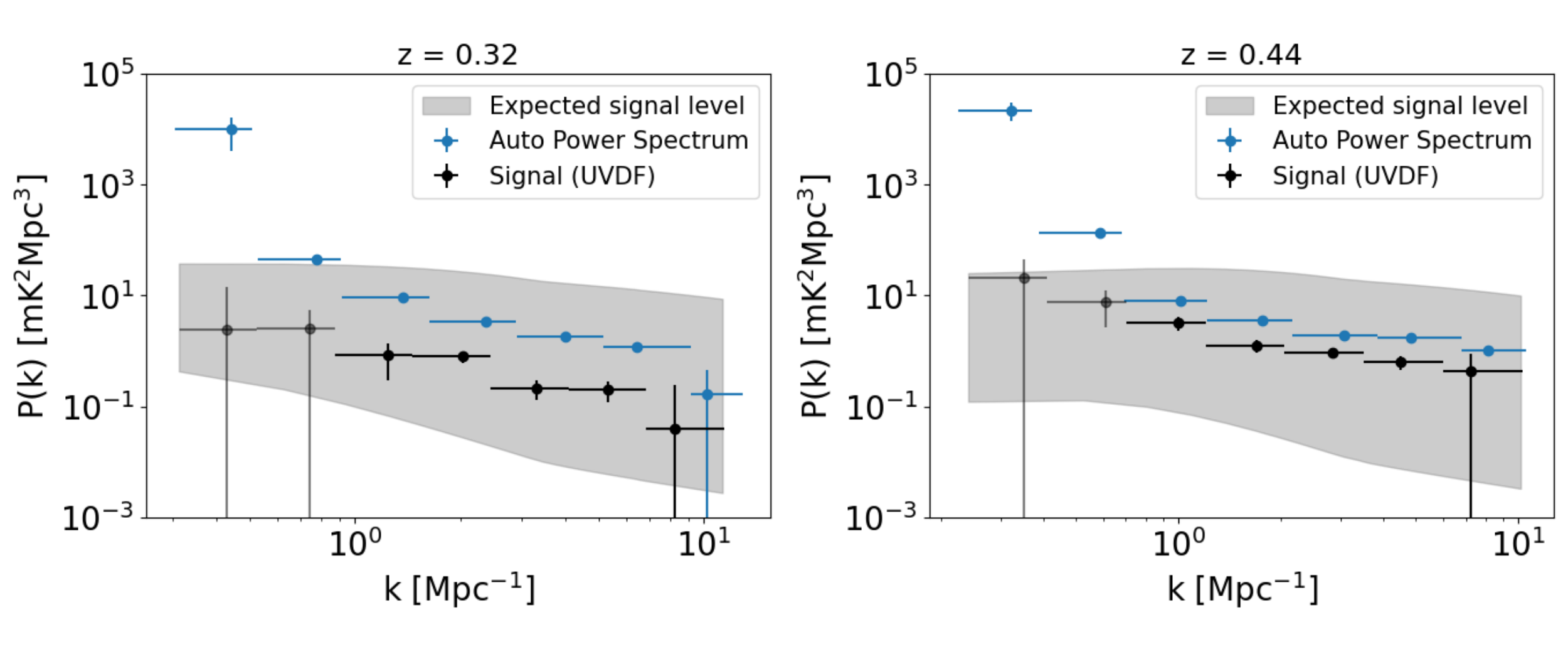}
\caption{{\bf Expected signal levels at $z=0.32$ and $z=0.44$.} The measurement with UVDF is denoted as ``Signal'' and the expected level of signal is calculated following \autoref{sec:model}. The first two $k$-bins, while demonstrating a detection, are believed to be less robust to systematics than the rest of the $k$-bins and, therefore, are denoted differently. The detection significance quoted in this work excludes the first two $k$-bins. The 1-d auto power spectrum in these plots is derived from the complete dataset with any baselines identified as contaminated by the BLF method excluded from the estimation and the noise bias subtracted. This auto power should be considered as an upper limit since it contains systematics that are further reduced in the cross-correlation.}
\label{1d_ps}
\end{figure*}

The measured power spectra have an amplitude of $\rm \sim 1\,mK^2Mpc^3$, decreasing to $\sim 0.1\,{\rm mK^2Mpc^3}$ at smaller scales. This is consistent with the combination of the expected shot noise level of $\sim 1\,{\rm mK^2Mpc^3}$ (e.g. \citealt{Chen_2021}) and heavy attenuation due to the Finger-of-God (FoG) effects at high $k_\parallel$. 
In \autoref{1d_ps}, we show a comparison of our measured signal amplitude with the expected model amplitude in the gray regions. We note that at the scales of our measurements, there are subtleties in the modelling of the HI power that have not been resolved, which we discuss briefly in \autoref{sec:model}. The expected signal level can vary over an order of magnitude, which is consistent with our measurement.

In \autoref{1d_ps}, we also present the 1-d auto power spectrum derived from the full dataset (noise bias removed), excluding baselines which are identified as contaminated by BLF -- with no additional flagging applied at a later stage. The auto power spectrum can be considered as a robust upper limit of the expected signal.

We use the 3-d power from our data to compute the variance of the fluctuation at a given scale $R$ to further quantify the amplitude of the HI fluctuations measured: 
\begin{equation}
\sigma_{\rm HI}^2 = \frac{3}{4\pi R^3}\,\frac{\int_{k_{\rm window}}\frac{{\rm d}^3k} {(2\pi)^3}W^2(kR)P(\bm{k})w(\bm{k})}{\int_{k_{\rm window}}\frac{{\rm d}^3k}{(2\pi)^3}W^2(kR)w(\bm{k})},
\label{eq:sigmahi}
\end{equation}
where the integration sums over all the $k$-pixels in the delay-transformed visibility data cube and $w(\bm{k})$ is the inverse covariance weight discussed before. $W(kR) = 3 j_1(kR)/(kR)$ is the Fourier transform of the spherical top-hat function with $j_1$ being the spherical Bessel function of the first kind.

As shown in \autoref{ps_table_BLF} and \autoref{ps_table}, our measurements are most sensitive to scales around 1.0 Mpc. We choose the UVDF results as they exhibit better S/N, and use the bootstrapping method to estimate $\sigma_{\rm HI}$ at $R= 1.0$ Mpc, by running 8000 realizations with each realization selecting $10^4$ k-pixels that are not flagged. In each realization, a value of $\sigma_{\rm HI}$ is calculated according to the $k-$pixels selected. This allows us to constrain the rms of HI density to be $\sigma_{\rm HI} = (0.44\pm 0.04)\,{\rm mK}$ at $z\sim0.32$ and $\sigma_{\rm HI} = (0.63\pm 0.03)\,{\rm mK}$ at $z\sim0.44$, at the scale of 1.0 Mpc. The results for $\sigma_{\rm HI}$ coherently sum over all $k$-points, and thus is an indicator for the overall statistical significance of the measurements.

\section{Pipeline validation}
In the following sections, we present the consistency checks to validate our analysis pipeline and the power spectrum results. Some of these tests refer to the UVDF pipeline but the conclusions can be extended to the BLF approach.

 \begin{figure*}[!ht]
    \centering
    \includegraphics[width=\linewidth]{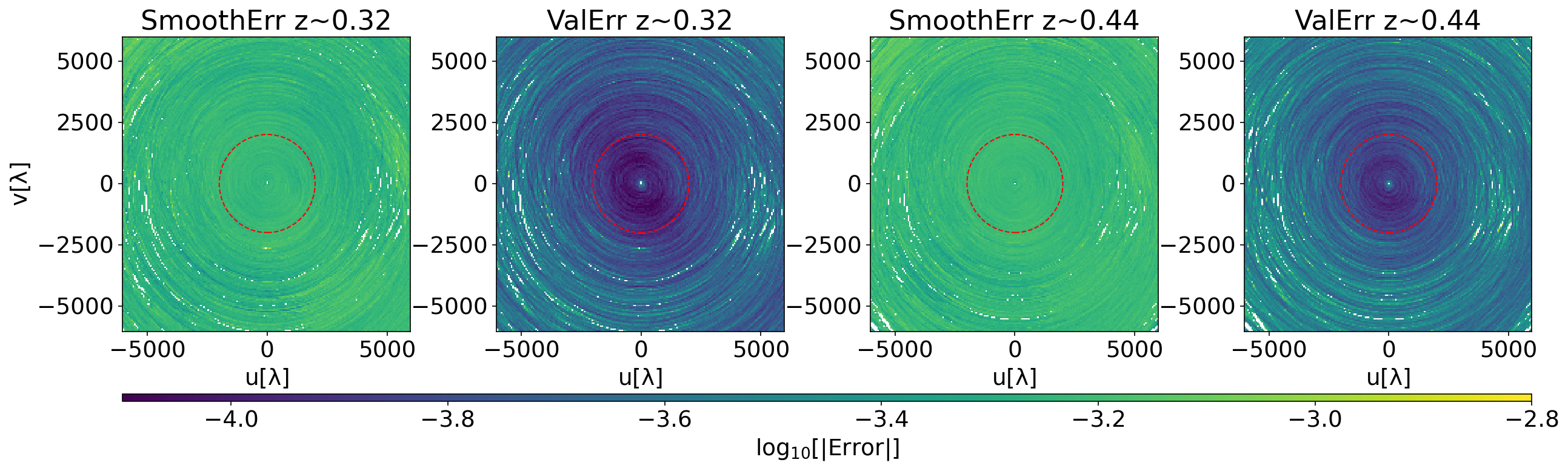}
    \caption{The gridded estimated bandpass error of the observation. The absolute values of the errors are taken and averaged across the frequency channels for visualization. Errors are estimated with two methods: ``ValErr'' and ``SmoothErr'' (discussed in \autoref{sec:gain}). The left panels show the results for the z=0.32 bin and the right for z=0.44. The red circles correspond to the $|u|<2000\,\lambda$ region where the visibilites are used for the power spectrum estimation.}
    \label{fig:grid_err}
\end{figure*}

\begin{figure*}
\centering
\includegraphics[trim={0 40 0 40},clip,width=1.0\textwidth]{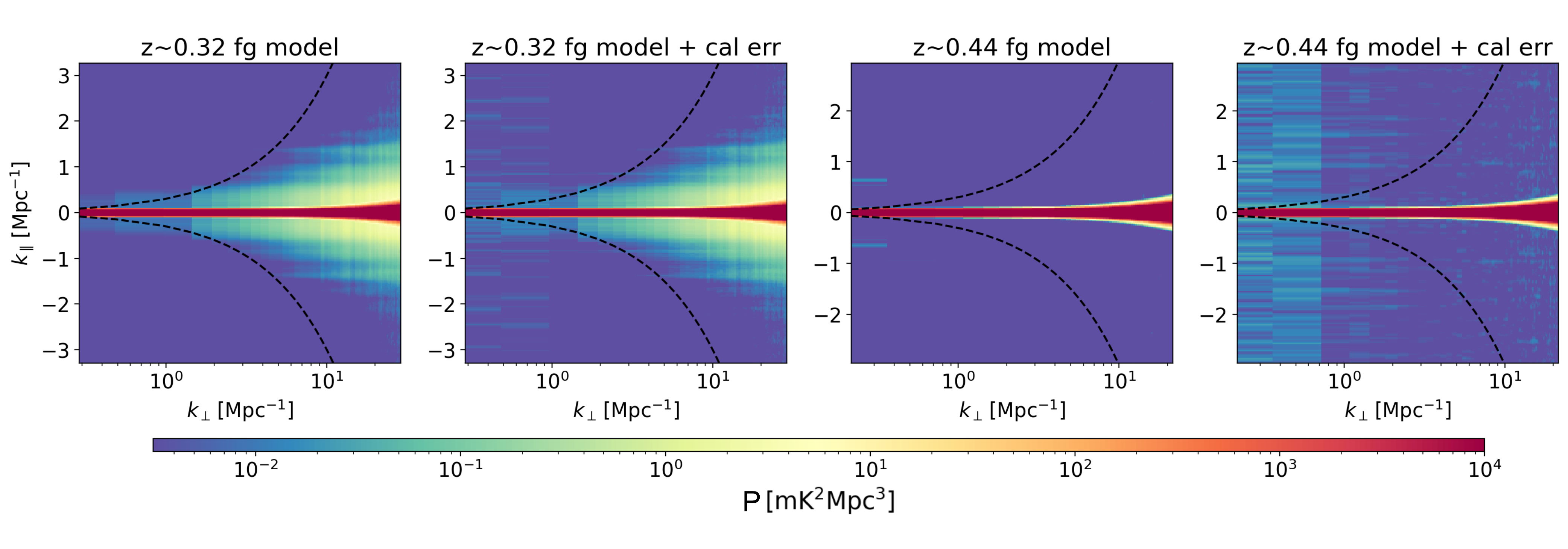}
\caption{{\bf Foreground scatter.} The cylindrical foreground power spectrum at $z=0.32$ without calibration error (left) and with calibration error (centre left), and at $z=0.44$ without calibration error (centre right) and with calibration error (right). Please note that the colour scale in this plot is different than that used in \autoref{fig:2d_PS_1} and \autoref{fig:2d_PS_2}.}
\label{fig:ps2dfg}
\end{figure*}

\subsection{Foreground scatter}
\label{subsec:fgscatter}
In our analysis, the assumption is made that foreground power is limited to lower Fourier modes and that the observation window ($k_\parallel > 0.3 k_\perp$) is not contaminated by foregrounds. To demonstrate the validity of this assumption, we estimate the foreground power spectrum using a model visibility dataset derived from a point source model. The point source model is generated by applying the \textsc{TCLEAN} task to the calibrated visibility data, and the same power spectrum pipeline used for the data is applied here, with identical processing steps (such as flagging, gridding, inpainting and thresholding). Foreground power can scatter to higher $k_\parallel$ modes primarily for two reasons: the inpainting of flagged channels and calibration errors across frequency. In this section, we examine the calibration errors and resulting foreground scatter.

The bandpass calibration errors, as discussed in \autoref{sec:gain}, are estimated using three different methods which are shown in \autoref{fig:gain_err}. The ``ValErr'' case assumes that the calibration error is random and averages down as the number of solution interval increases, corresponding to an optimistic case. The ``SmoothErr'' case assumes that deviation from the smooth spline corresponds to the calibration errors, which gives a moderate estimation of error. The ``PolyErr'' case, which is derived by fitting the bandpass solution with a tenth-order polynomial, and using the residual of the fitting as the calibration error. In this case, the structure of the bandpass in small frequency scales, which is likely to be real structures due to standing waves, is considered to be calibration errors, corresponding to a pessimistic estimation of calibration errors. {We emphasise that, these errors are directly estimated from the gain solutions, which we then use to propagate to a sky model as discussed below.}

The calibration errors at each dish and feed are propagated into the foreground model visibility, which is then flagged, inpainted, and gridded to produce the foreground power spectrum.
 The bandpass error at each baseline $\delta g_{ij}$ is a multiplication of the error at each feed so that
 \begin{equation}
     \delta g_{ij} = (1+\delta g_i) (1+\delta g_j) -1,
 \end{equation}
 where $i,j$ denotes the two feeds of each visibility data point.
 The fraction $\delta g_{ij}+1$ at baseline level is then gridded to each $(uv)$ cell using the same visibility gridding procedure as done for the power spectrum estimation.
 The results are shown in Figure \ref{fig:grid_err}. For simplicity, we only show the results for ``ValErr'' and ``SmoothErr'', as ``PolyErr'' and ``SmoothErr'' give similar results. For both redshift bins, we find the overall bandpass error to be $< 10^{-3}$ on the $uv$ pixels, sufficient for containing foreground leakage \citep{2016MNRAS.461.3135B}. We also find that, overall, the ``SmoothErr'' case gives more delay scatter.

\begin{figure*}
    \centering
    \includegraphics[trim={0 40 0 30},clip,width=1.0\linewidth]{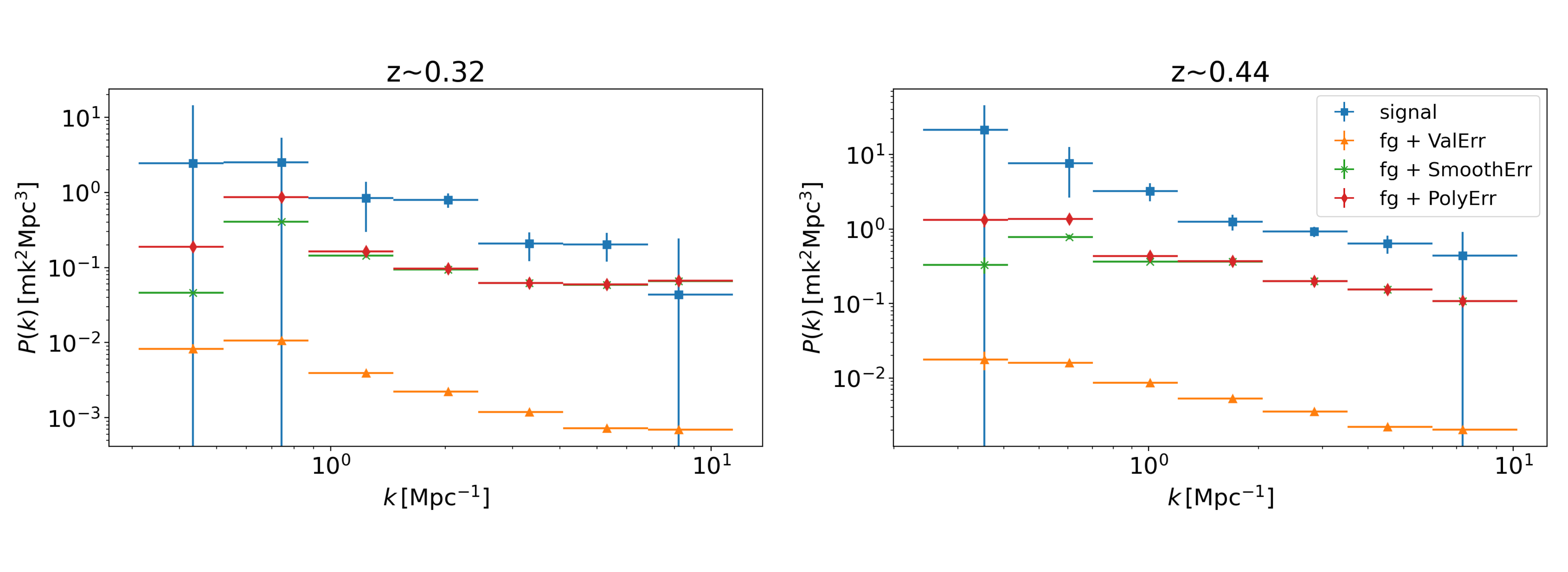}
    \caption{The 1-d foreground power spectrum at $z=0.32$ and at $z=0.44$. Estimated calibration errors are applied to the model foreground visibility, with three different scenarios corresponding to the optimistic (``fg + opt''), moderate (``fg + mod''), and pessimistic (``fg + pes'') estimations described in \autoref{subsec:fgscatter}. The measured signal is also shown for comparison (`signal').}
    \label{fig:ps1dfg}
\end{figure*}

We then combine the gridded bandpass errors with the point sources visibility model. From this we calculate the power spectrum following the same procedure as before. In Fig. \ref{fig:ps2dfg}, we show the cylindrical power spectrum of the foregrounds for both redshift bins. Only the ``ValErr'' case is shown here for illustration. The foreground power is well suppressed below the horizon line for the $z=0.44$ bin, while for the $z=0.32$ bin, there is some scatter in delay, especially for the long baselines. This is due to the fact that the $z=0.32$ bin has a sub-band around 1090MHz that is heavily flagged, while the $z=0.44$ bin is quite evenly sampled. When the calibration error is considered, a general rise in power is observed in all $k$ modes. Still, the scatter of foreground power above the horizon is insignificant, which is orders of magnitude lower than the measured signal. This can be verified in the 1-d power spectrum as shown in \autoref{fig:ps1dfg}.

We can see that when the optimistic calibration error ``ValErr' is applied to the foreground power, the foreground scatter is negligible. This suggests that the integration time on the calibrator scans is sufficient to produce calibration solutions with high signal-to-noise ratios. On the other hand, when ``SmoothErr'' is applied to the foreground model, the foreground scatter is amplified by $\sim$2 orders of magnitude. For the $z\sim 0.44$ case, the estimated foreground power is consistently lower than the signal by at least a factor of 5. For $z\sim 0.32$, however, the foreground scatter is more severe due to the higher flagging fraction of the data. The flagged visibilities are inpainted using the nearest unflagged channel, causing the foreground scatter. This results in the first two $k$-bins as well as the last $k$-bin showing foreground contamination being closer to the signal level. We note that our power spectrum estimator also gives enlarged error bars in these $k$-bins, indicating the existence of extra scatter in the data. This does not affect our detection significance, as these $k$-bins do not have enough signal-to-noise ratio to achieve detection. For the scales at $1\,{\rm Mpc^{-1}}< k < 5\,{\rm Mpc^{-1}} $, the detection is robust against a pessimistic estimation of foreground scatter. {As discussed in Section \ref{sec:gain}, the deviations of bandpass away from a smooth solution are caused by two effects, standing waves and receiver dips. The bandpass solutions correct for these effects, which we quantify conservatively as calibration errors in the case of ``SmoothErr'' and ``PolyErr''. Therefore, the results shown in \autoref{fig:ps1dfg} are overestimations of foreground scatter.}

\subsection{Signal extraction checks}\label{signal_recovery}
\begin{figure*}
    \centering
    \includegraphics[width=0.9\linewidth]{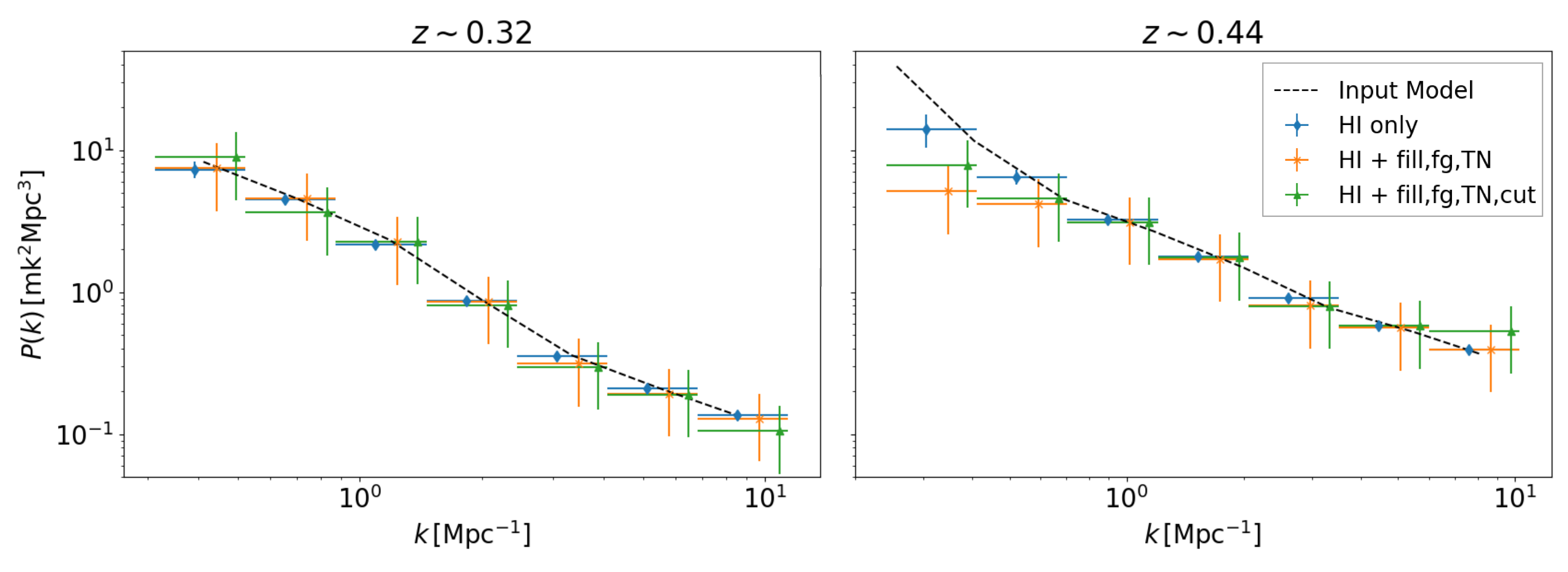}
    \caption{The 1-d power spectrum of the simulated HI visibilities averaged across 100 realizations at $z=0.32$ (left panel) and $z=0.44$ (right panel). The power spectra of the simulated HI without flags and inpainting (`HI only'), with model foregrounds and thermal noise including flags and inpainting (`HI + fill,fg,TN'), and further with the removal of $k$-pixels and inverse variance weighting identical to the measurement (`HI + fill,fg,TN,cut'). The model power spectrum is shown as the black dashed line. The centres of the $k$-bins are misplaced for better visualization. All cases use the same measurement window ($k_\parallel = 0.3 k_\perp$).}
    \label{fig:ps1dhisim_alt}
\end{figure*}

In this section, we discuss the recovery of a known HI signal in the presence of foreground and thermal noise with our power spectrum analysis pipeline. We follow the methodology described in \citet{Paul_2021} to generate HI visibilities on a delay-transformed $uv$ grid, incorporating the best-fit power spectrum (see \autoref{modelfit}). The foreground component is derived from the field's point source model, while thermal noise is added as Gaussian random values for each $uv$ point. The simulated visibility data is passed through the identical steps as the real data such as inpainting, gridding and power spectrum estimation pipeline. In \autoref{fig:ps1dhisim_alt}, we compare the 1-d results between (a) only HI component without flags and inpainting in blue, (b) HI with model foreground and thermal noise including the data flags and inpainting in yellow, and (c) further adding steps such as the removal of $\mathbf{k}$ pixels from the UVDF method and inverse variance weighting in green. \autoref{fig:ps1dhisim_alt} establishes that our power spectrum pipeline is able to recover the input HI signal and the signal loss caused by inpainting and thresholding is negligible compared to the measurement error.

\subsection{Jackknife tests}
\begin{figure*}
    \centering
    \includegraphics[width=\linewidth]{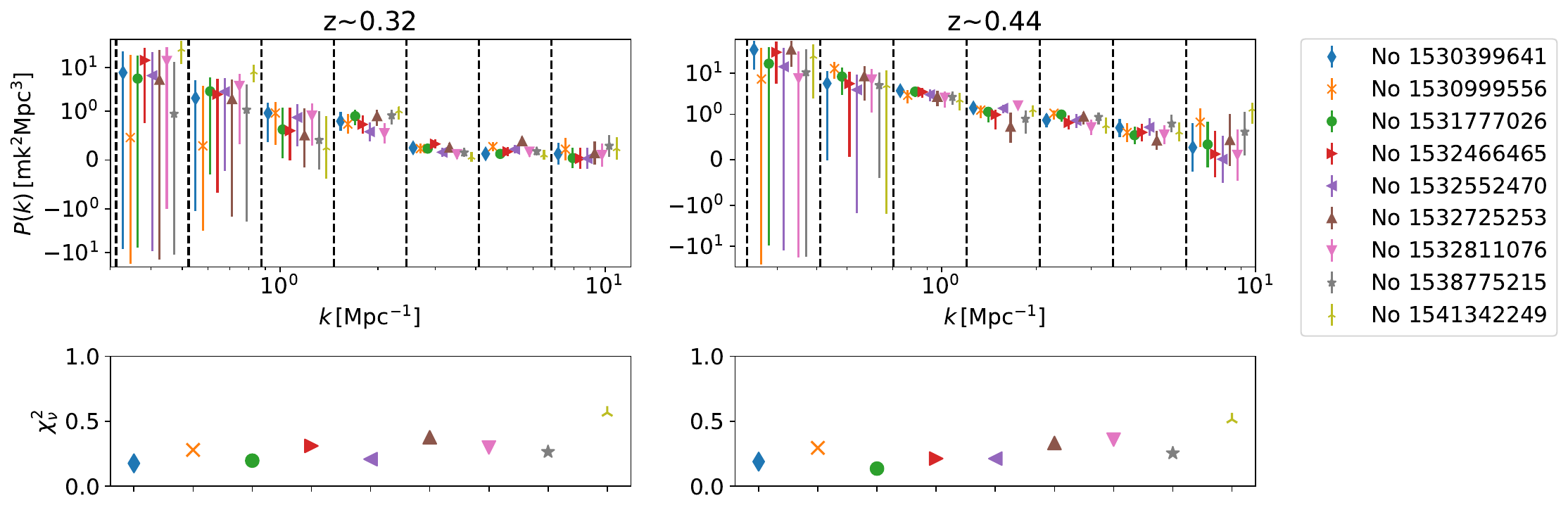}
    \caption{{\bf Jackknife test.} The 1-d power spectrum of the measured visibility dataset when one observation block is excluded. The black dashed line shows the boundary between different $k$-bins. The $k$-coordinates of different sub-samples are placed evenly inside the $k$-bins for visualization. Each sub-sample has one observation block excluded from the visibility data. The id of the block excluded is shown on the label. {The bottom panels show the reduced chi-square, comparing each jackknife realisation to the overall measurements.} }
    \label{fig:psjack}
\end{figure*}

To test possible biases and the statistical consistency of our results, we perform jackknife tests where we systematically exclude one observation block at a time. We apply the exact same processing steps to estimate the power spectrum of the HI signal. The jackknife results in the 1-d power spectrum demonstrate coherence across all observational blocks as seen in \autoref{fig:psjack}. We use the identical $k$-mask derived from the $5$-$\sigma$ flagging of the complete dataset in UVDF. Notably, all sub-samples exhibit coherence with each other across all $k$-bins. Fluctuations are consistent with thermal noise expectations, showing no indication of non-negligible systematics.

\subsection{Null tests}
\label{validation-null}
We also perform a null test by cross-correlating the data of the two frequency sub-bands used in the detection. The cosmological signal has no correlation between the two redshift bins, so any non-zero correlation would be an indication of residual systematics. Each of the individual sub-bands is split into two time blocks of even and odd scans. When calculating the cross-power, only $k$-points that are above the $k_\parallel = 0.3 k_\perp$ wedge and not flagged by the $5-\sigma$ criterion are used. 

\begin{figure*}
\centering
\includegraphics[trim={0 30 0 15},clip,width=\textwidth]{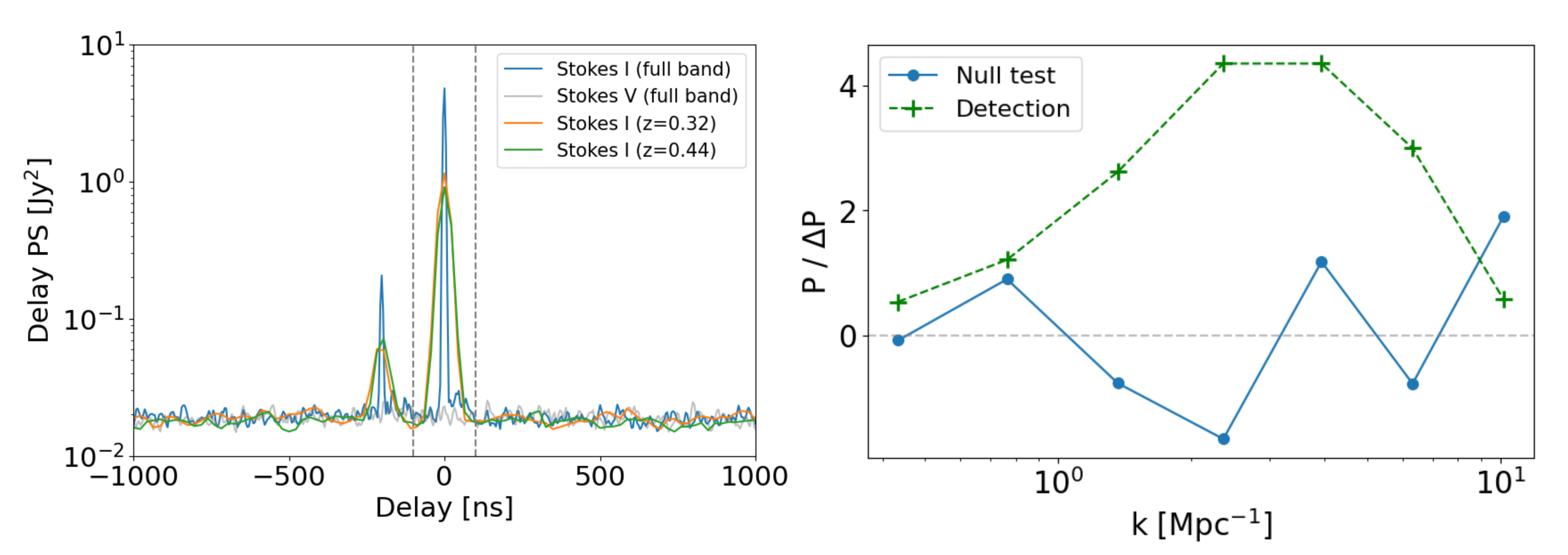}
\caption{{\bf Evidence of broadband systematics and Null test.} {\bf Left:} Delay power spectra for the same antenna pair and scan as \autoref{fig:bl_flagging_1}. The Stokes~I spectra are shown separately for the two analysis bands, $z=0.32$ and $z=0.44$, and for the full accessible frequency range $952{\text -}1170$ MHz (labeled as `full band'). The Stokes~V full band spectrum is shown as a noise reference. The vertical dashed lines mark $|\tau|=100$ ns, enclosing the foreground-dominated delay region. The excess features appear at the same physical delays in the two individual spectral windows and in the full-band transform, demonstrating that the systematic is broadband and coherent across the analysis bands. {\bf Right:} The ratio of the estimated cross power (odd × even) to its corresponding $1\sigma$ error is shown for two different cases. The `Detection' case corresponds to the results presented in this paper using UVDF, where $P/\Delta P$ is averaged over $z=0.32$ and $z=0.44$. The `Null test' case represents the average cross power to error ratio where the odd and even scans belong to different $z$-bins. The $k$-values are computed at $1078$ MHz. The absence of signal in the `Null test' (Right), despite the broadband nature of the contaminant (Left), confirms that our flagging strategy successfully isolates and removes these systematics from the final measurement.}
\label{null_test}
\end{figure*}

Before examining the cross-correlation, it is crucial to establish the spectral properties of the residual systematic contamination. In the left panel of \autoref{null_test}, we calculate the delay power spectrum for the same contaminated baseline and scan shown in \autoref{fig:bl_flagging_1} over the full accessible frequency range of $952{\text -}1170$ MHz. The presence of a distinct contamination spike at the same physical delay across this wider bandwidth demonstrates that the systematics is indeed broadband, exhibiting coherence across both redshift bins. Quantitatively, comparing the two individual-band Stokes~I delay spectra over the non-foreground delay range $100<|\tau|<1000$ ns gives a strong Pearson correlation coefficient of $r_\tau=0.93$. This provides direct evidence that the contamination is coherent across the two spectral windows.

This broadband coherence makes the null test a highly diagnostic tool. If our foreground avoidance and flagging strategies were failing to mitigate this systematic, its spectral coherence would inevitably cause it to leak into the final measurement, producing a strong, false detection when cross-correlating the two frequency bands. As shown in right panel of \autoref{null_test}, we detect no significant correlation in these tests which shows that no detectable frequency correlated contaminations remain in the Fourier window used in the analysis. The null test suggests that the observation window we choose does not have any sizeable foreground leakage, and residual systematics have been largely mitigated.

\section{Model fitting}
\label{modelfit}
In this section, we use the obtained measurement of the HI power spectrum with UVDF to constrain the astrophysics of the HI galaxies and the HI mass function (HIMF) following \citet{Chen_2021} (also see \citealt{2023arXiv230509720P}). 

\subsection{Power spectra}
The power spectrum is modelled as a combination of a 2-halo and 1-halo term as well as a scale-independent shot noise term. We further include redshift space distortions, e.g. the Kaiser term \citep{Kaiser_1987} which increases the amplitude of the power spectrum on large scales and the ``fingers-of-god" term \citep{Jackson_1971} which smooths out fluctuations on small scales. The theoretical power spectrum is averaged in the same way as the observed one in order to produce the 1-d power spectrum.

We model the HI power spectrum as a combination of a 2-halo, 1-halo, and shot noise term \citep{Cooray_2002,2010MNRAS.404..876W,2019MNRAS.484.1007W, Chen_2021}:
\begin{linenomath*}
\begin{equation}
    P_{\rm D}(k_\perp,k_\parallel) = P_{\rm 2h}(k_\perp,k_\parallel) + P_{\rm 1h}(k_\perp,k_\parallel) + P_{\rm SN}(k_\parallel).
\end{equation}
\end{linenomath*}
The 2-halo term can be modelled as
\begin{equation}
\begin{split}
P_{\rm 2h}(k,k_\parallel) =& C_{\rm HI}^2 \bigg[\int{\rm d}m\,n(m) b(m)\langle M_{\rm HI}(m) \rangle u_{\rm HI}(k|m) \bigg]^2\\
&(1+\frac{fk_\parallel^2}{b_{\rm HI}^0k^2})^2 \frac{P_{m}(k)}{{1+(k_\parallel\sigma_p)^2}/2},
\end{split}
\end{equation}
where $m$ integrates over the halo mass range, $C_{\rm HI}$ is the conversion factor from HI mass density to temperature, $n(m)$ is the halo mass function \citep{Tinker_2008}, $b(m)$ is the linear halo bias \citep{Tinker_2010}, $\langle M_{\rm HI}(m) \rangle$ is the HI-halo mass relation, $u_{\rm HI}(k|m) = \mathcal{F}(\rho_{\rm HI}(r|m))$ is the normalised Fourier-transform of the density profile $\rho_{\rm HI}(r|m)$, $k=\sqrt{k_\perp^2+k^2_\parallel}$ and $b_{\rm HI}^0$ is the linear HI bias. We include redshift space distortions in our analysis, taking into account both the Kaiser effect \citep{Kaiser_1987} and the FoG effect \citep{Jackson_1971} which affects the small line of sight scales (large $k_\parallel$) we are observing. For these terms, $f$ is the growth rate and $\sigma_p =\sigma_v(1+z)/(Hf) $ where $\sigma_v$ is the velocity dispersion and $H$ is the Hubble parameter. Here, we simplify the modelling of the FoG effects by assuming an averaged velocity dispersion $\sigma_v$, which can be related to the velocity width of the emission line profile of the stacked HI galaxies.
$P_{\rm m}(k)$ is the matter power spectrum calculated using halofit \citep{Takahashi_2012}. 
 
 The linear HI bias is calculated as
 \begin{linenomath*}
\begin{equation}
    b_{\rm HI}^0 = \frac{\int{\rm d}m\,n(m) b(m)\langle M_{\rm HI}(m) \rangle}{\int{\rm d}n\,n(m) \langle M_{\rm HI}(m) \rangle},
\end{equation}
\end{linenomath*}
while the 1-halo term can be modelled as
\begin{equation}
\begin{split}
    P_{\rm 1h}(k,k_\parallel) =C_{\rm HI}^2 \int {d}m \,n(m)\langle M_{\rm HI}(m)\rangle ^2 \\
    \times u_{\rm HI}(k|m)\frac{1}{{1+(k_\parallel\sigma_p)^2}/2} .
\end{split}    
\end{equation}

The shot noise term $P_{\rm SN}(k_\parallel)$ can be written as
\begin{linenomath*}
\begin{equation}
    P_{\rm SN}(k_\parallel) = \frac{P_{\rm SN}^0 }{1+(k_\parallel\sigma_p)^2/2}.
\end{equation}
\end{linenomath*}
Note that, the shot noise in the HI power spectrum is also attenuated by the FoG effect. This is due to the fact that the HI sources are not treated as point sources on the sky but as intensity maps. The emission line width, corresponding to the velocity dispersion of the sources, breaks the point-source assumption along the line of sight, smearing the overall power spectrum along the $k_\parallel$ direction. 

We further parameterize:
\begin{equation}
\begin{split}
        \langle M_{\rm HI}(m)\rangle = A_{\rm HI}\big(\frac{m}{M_0}\big)^\beta,\,\\ \rho_{\rm HI}(r|m) = \rho_0 \big(\frac{r}{r_0}\big)^\gamma {\rm exp}[-ar/r_0],
\end{split}
\end{equation}
where $M_0 = 10^{10} M_\odot {\rm h}^{-1}$, $r$ is the distance relative to the halo centre, and $r_0$ is the characteristic scale of the dark matter halo, calculated using the concentration-mass relation \citep{Maccio_2006}. The parameter $a$ is fixed by imposing that at $m=10^{12} M_\odot {\rm h}^{-1}$ the cut-off scale is $r_0/a = 1 {\rm Mpc \,h^{-1}}$ \citep{Spinelli_2019}. We further fix the power law index for the velocity dispersion as in \cite{Villaescusa-Navarro_2018}, which leaves us with a 5-parameter model,
\begin{linenomath*}
\begin{equation}
    \theta = \{A_{\rm HI}, \beta, \gamma, \sigma_v, P_{\rm SN}^0  \}.
\end{equation}
\end{linenomath*}

\subsection{HI mass function}
From the inferred halo model parameters, the overall HI density fraction of the Universe, $\Omega_{\rm HI}$, can be calculated as
\begin{linenomath*}
\begin{equation}
    \Omega_{\rm HI} = \int{\rm d}m\,n(m) \langle M_{\rm HI}(m) \rangle /\rho_c^0,
\label{eq:omegahihalo}
\end{equation}
\end{linenomath*}
where $\rho_c^0$ is the critical density of the Universe at $z=0$.

Note that, in the above modelling, the real-space shot noise $P_{\rm SN}^0$ is an independent parameter, which is agnostic to the choice of the halo model and consequently unrelated to the over HI density $\Omega_{\rm HI}$. This means that to fit a 1-d power spectrum as a combination of 2-halo, 1-halo and shot noise term, one can have an arbitrarily small 2-halo and 1-halo contribution, as the shot noise term can dominate the power spectrum fitting. To resolve this issue, we need a formalism to relate the HI shot noise to the overall HI density.

The relation between the HI shot noise and the HI density can be established using the HI mass function (HIMF). The HIMF is parameterized with a Schechter function \citep{Schechter_1976}:
\begin{linenomath*}
\begin{equation}
    \phi_{\rm HI}(M_{\rm HI})\equiv \frac{{\rm d}n_{\rm HI}}{{\rm dlog}M_{\rm HI}} = {\rm ln}10\,\phi_* \big(\frac{M_{\rm HI}}{M_*}\big)^{\alpha +1}e^{-\frac{M_{\rm HI}}{M_*}},
\end{equation}
\end{linenomath*}
where $\phi_*$ is the overall amplitude of the HIMF, $\alpha$ is the slope of the HIMF before the ``knee mass'' $M_*$.

The HIMF can be related to the number density, $n_{\rm HI}$, HI mass density and shot noise as \citep{Chen_2021}:
\begin{equation}
    n_{\rm HI} = \int_{M_{\rm min}} {\rm dlog}M_{\rm HI}\,\phi_{\rm HI}(M_{\rm HI})
\label{eq:nhi}
\end{equation}
\begin{equation}
    \Omega_{\rm HI} = \int {\rm dlog}M_{\rm HI}\,\phi_{\rm HI}(M_{\rm HI}) M_{\rm HI}/\rho_c
\label{eq:omegahi}
\end{equation}
\begin{equation}
    P_{\rm SN}^0 = (C_{\rm HI}\rho_c \Omega_{\rm HI})^2 \frac{\int {\rm dlog}M_{\rm HI}\,\phi_{\rm HI}(M_{\rm HI})M_{\rm HI}^2}{\big(\int {\rm dlog}M_{\rm HI}\,\phi_{\rm HI}(M_{\rm HI})M_{\rm HI}\big)^2}
\label{eq:psn}
\end{equation}

The Schechter function has three parameters, constrained by the number density, mass density and the shot noise of HI galaxies. Using the measured HI power spectrum, we are able to constrain the shot noise $P_{\rm SN}^0$ and, through the halo model, constrain $\Omega_{\rm HI}$. However, the technique of intensity mapping is not sensitive to the number density of HI galaxies. Therefore, the HIMF parameters are unlikely to be fully constrained by our model fitting. Instead, we intend to use the HIMF parameters with physically driven priors, to exclude part of the parameter space with extremely high values of $P_{\rm SN}^0$ and low values of $\Omega_{\rm HI}$.

For a given set of halo model parameters, we construct the parameter set by adding two HIMF parameters, $M_*$ and $\alpha$, so that
\begin{linenomath*}
\begin{equation}
    \theta = \{A_{\rm HI}, \beta, \gamma, \sigma_v, M_*, \alpha \}.
\end{equation}
\end{linenomath*}
From \autoref{eq:omegahihalo}, we can compute the HI density using the halo model parameters, and then use \autoref{eq:omegahi} to determine the remaining HIMF parameter $\phi_*$. Therefore, the final parameter set has 6 free parameters.

\begingroup

\begin{table}
    \centering
    \renewcommand{\arraystretch}{1.2}
    \begin{tabular}{|c|c|c|c|}
        \specialrule{0.25mm}{0pt}{0pt}
        ${\rm log}_{10}[A_{\rm HI}/M_\odot]$ & $\beta$ & $\gamma$ & $\sigma_v\,$[km/s]  \\\hline
        [0,14] & [0,10]  & [1,3] & [0,500] \\
        \specialrule{0.25mm}{0pt}{0pt}
        ${\rm log}_{10}[M_*/M_\odot]$ & $\alpha$ & $\Omega_{\rm HI}^{z=0.32}\times 10^{3}$  & $\Omega_{\rm HI}^{z=0.44}\times 10^{3}$  \\\hline
        [6,13] & [-5,0] & $\mathcal{N}(0.50,0.18)$ & $\mathcal{N}(0.77,0.26)$ \\
        \specialrule{0.25mm}{0pt}{0pt}
    \end{tabular}
    \caption{The priors adopted in the parameter fitting of this work. All model parameters have flat priors. The NP case in the figures does not have any priors on $\Omega_{\rm HI}$, 
    while the WP case uses the Gaussian priors listed in the table.}
    \label{table:prior}
\end{table}
\endgroup

\subsection{Constraints}
After choosing the halo model parameters, we compute the analytical power spectrum using \textsc{halomod} \citep{Murray_2020} and average across the k-range probed as shown in \autoref{fig:2d_PS_1}a and \autoref{fig:2d_PS_2}a. We construct a Gaussian likelihood from the results reported and use \textsc{nautilus} \citep{nautilus} to perform Bayesian inference using importance nested sampling. Flat, wide priors are imposed as listed in \autoref{table:prior}. For $\gamma$, we impose $1<\gamma<3$ since values outside this range lead to divergence in the density profile integration. We impose a flat prior on the velocity dispersion $0<\sigma_v<500\,{\rm km\,s^{-1}}$, since the velocity dispersion is related to the HI mass of the galaxies through the Tully-Fisher relation (e.g. \citealt{2021MNRAS.508.1195P}). Velocity dispersion larger than $500\,{\rm km\,s^{-1}}$ implies unphysical massive HI galaxies.

As we show later, the parameter fitting does not converge for all parameters due to parameter degeneracy and lack of information on the $k_\parallel$ dependence of the power spectrum. We therefore consider another case where we impose a prior on $\Omega_{\rm HI}$ with $\Omega_{\rm HI} = (0.50\pm0.18) \times 10^{-3}$ at $z\sim 0.32$ from stacking \citep{Rhee_2018} and $\Omega_{\rm HI} = (0.77\pm0.26) \times 10^{-3}$ at $z\sim 0.44$ from Damped Lyman Alpha (DLA) systems \citep{Rao_2017}.

\begin{figure*}
\centering
\includegraphics[width=0.9\textwidth]{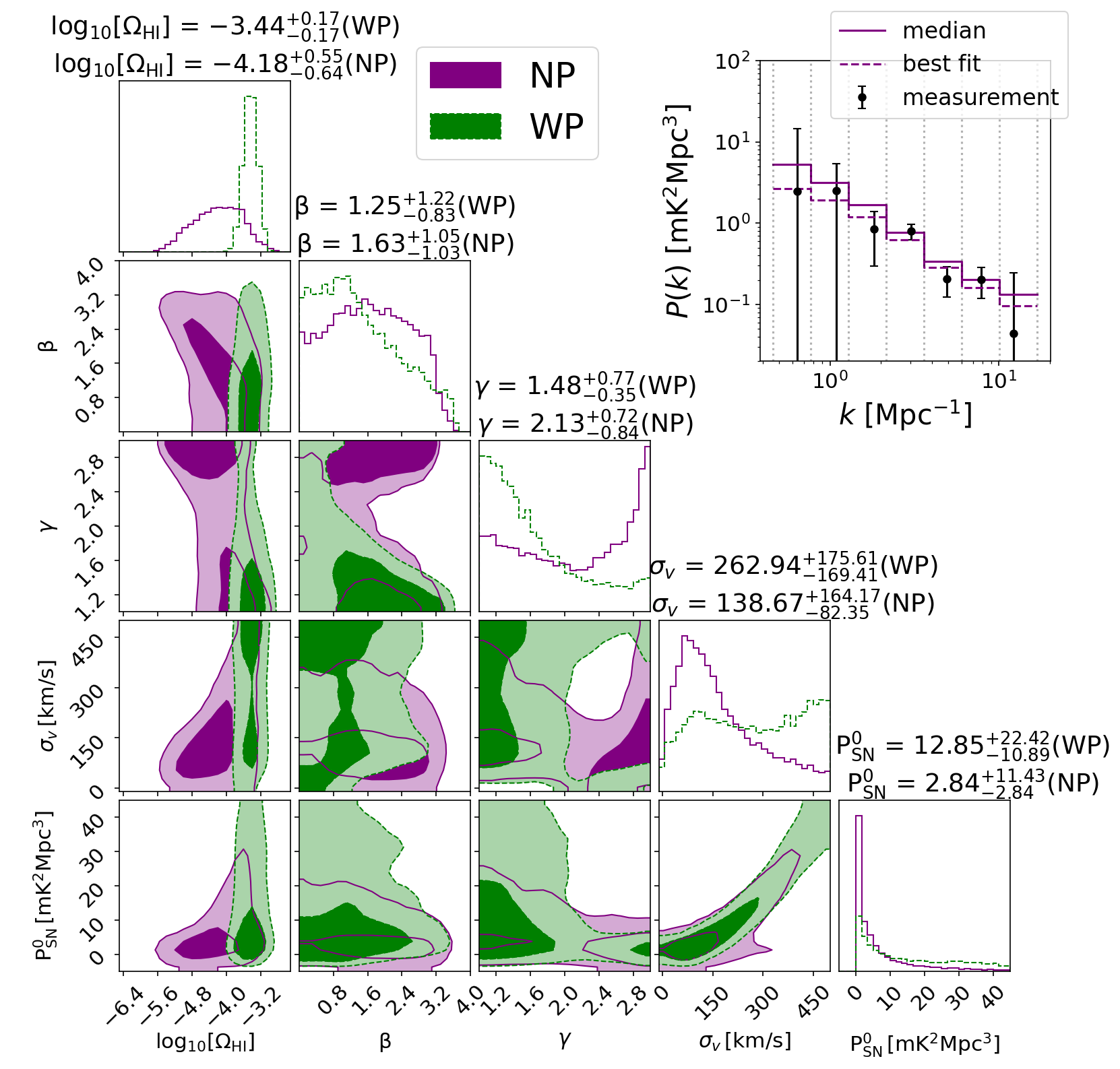}
\caption{{\bf Model fitting results for $z\sim0.32$.} The 2-d contour plots show the posterior distribution of model parameters for power spectrum fitting. The darker region shows the $1\sigma$ confidence interval of the posterior and the lighter region shows the $2\sigma$ confidence interval. The upper lines of the titles of the 1-d histograms show the median and the $1\sigma$ confidence level for each parameter for the model fitting with the $\Omega_{\rm HI}$ prior (``WP''). The lower lines of the titles of the 1-d histograms show the results without the $\Omega_{\rm HI}$ prior (``NP''). The 2-d posterior distribution is smoothed with a Gaussian kernel of $0.5$ grid width for better visualization. In the upper right zoom-in panel, the fitting result of the power spectrum measurement with no prior is shown. The vertical dash line shows the lower and upper limit for each k-bin. }
\label{contour1}
\end{figure*}

\begin{figure*}
\centering
\includegraphics[width=0.9\textwidth]{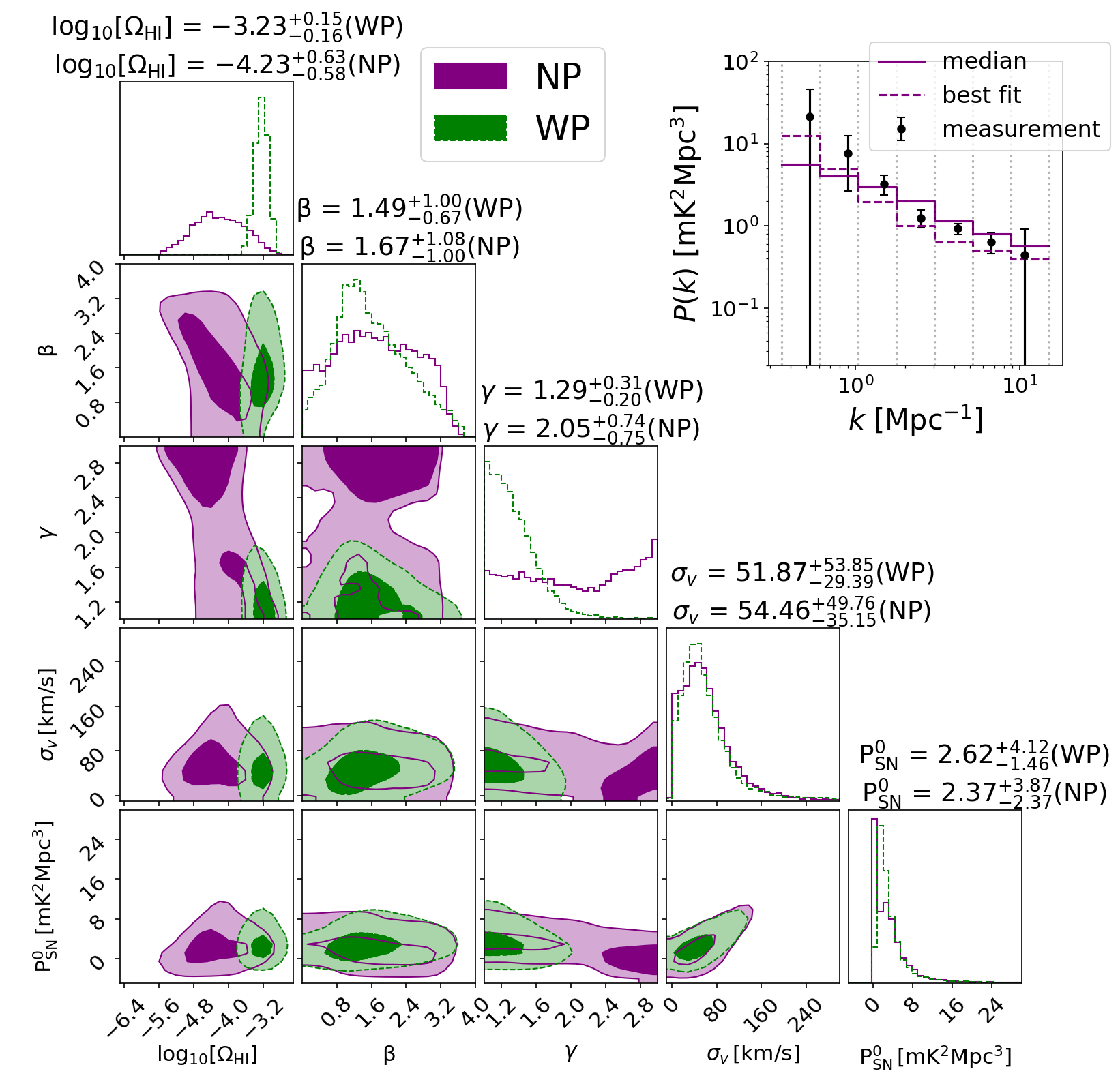}
\caption{{\bf Model fitting results for $z\sim0.44$.}The 2-d contour plots show posterior distribution of model parameters for power spectrum fitting. The darker region shows the $1\sigma$ confidence interval of the posterior and the lighter region shows the $2\sigma$ confidence interval. The upper lines of the titles of the 1-d histograms show the median and the $1\sigma$ confidence level for each parameter for the model fitting with the $\Omega_{\rm HI}$ prior (``WP''). The lower lines of the titles of the 1-d histograms show the results without the $\Omega_{\rm HI}$ prior (``NP''). The 2-d posterior distribution is smoothed with a Gaussian kernel of $0.5$ grid width for better visualization. In the upper right zoom-in panel, the fitting result of the power spectrum measurement with no prior is shown. The vertical dash line shows the lower and upper limit for each k-bin. }
\label{contour2}
\end{figure*}

The HOD amplitude $A_{\rm HI}$ is specific to the parameterization, and its physical meaning can be hard to interpret. Therefore, in each sample in the posterior, we use $\Omega_{\rm HI}$ to present the results instead of $A_{\rm HI}$.
We first convert the posterior on HIMF parameters back to the shot noise, and show the fitting results in \autoref{contour1} for $z\sim0.32$ and \autoref{contour2} for $z\sim0.44$. The green regions show the posterior distribution for the model parameters with the $\Omega_{\rm HI}$ priors, while the purple region shows the distribution without the prior. From the posterior, one can see that the small number of $k$-bins and the relatively low signal-to-noise ratio lead to divergence in the MCMC fitting, with most of the parameters unconstrained. The constraints on $\Omega_{\rm HI}$ are noticeably loosened and HI density as low as $\Omega_{\rm HI}\sim 10^{-5}$ if no prior is imposed. This is expected, since the small inner-halo scales do not contain enough information to constrain the overall amplitude which is on the large scales. The density profile parameter $\gamma$ is also poorly constrained, because the small scale power spectrum is dominated by the effects of the shot noise and the velocity dispersion.

Comparing the results at $z\sim 0.32$ and $z\sim 0.44$, one can see that both redshift bins show tangible evidence that the velocity dispersion $\sigma_v$ and the HI shot noise can be jointly constrained. For $z\sim 0.32$, the posterior is reaching the flat prior $\sigma_v<500\,{\rm km/s}$ we impose. We verify that relaxing this prior simply leads to the posterior stretching to higher, more unphysical values of $\sigma_v$. The large tail of the $\sigma_v$ distribution also drives the shot noise $P_{\rm SN}^0$ to be larger, since these two parameters are degenerate.

The $z\sim 0.44$ redshift bin has higher signal-to-noise ratio and the results give a reasonable constraint on the velocity dispersion $\sigma_v = 54.46^{+49.76}_{-35.15} \,{\rm km/s}$ and on shot noise ${\rm P_{SN}^0} = 2.37^{+3.87}_{-2.37} \,{\rm mK^2Mpc^3}$, although the problem of the degeneracy between $\sigma_v$ and $P_{\rm SN}^0$ persists. This is again expected given our current measurements. The results are robust against the choice of priors, as one can see that $\sigma_v$ no longer occupies the entire prior volume. The cylindrical power spectrum for the DEEP2 data is dominated by thermal noise, and therefore we need to average the $k$-pixels down to 1-d power spectrum in order to constrain our model. Binning the cylindrical power spectrum into 1-d power spectrum means that we lose the information along the $k_\parallel$ direction, which entails the effects of the $\sigma_v$ parameter. In the future with more data resulting in higher signal-to-noise ratio, binning the $k$-points into $k_\perp-k_\parallel$ grids will help preserve the $k_\parallel$ dependence of the FoG effect, breaking the degeneracy between $\sigma_v$ and the rest of the parameter set.

Comparing the results with and without the $\Omega_{\rm HI}$ prior, we find that the fitting converges on the $\Omega_{\rm HI}$ parameter as expected, but shows no visible improvement for shot noise and velocity dispersion. For $z\sim 0.44$ with better signal-to-noise ratio, we start to be able to constrain the density profile parameter $\gamma$ and the slope of HI HOD $\beta$, which leads to better constraints on the HI bias at small scales. The results suggest that at the 1-d power spectrum level, the signal at the scales of the measurement is dominated by the 1-halo term and the shot noise, which are less correlated with the overall HI amplitude compared to the large scales dominated by the 2-halo term.

Finally, we note that without the $\Omega_{\rm HI}$ prior, the halo model fitting produces broad constraints on $\Omega_{\rm HI}$. Comparing the posterior of the no prior case with the prior, one can see that they are consistent with each other. However, the priors for both the $z\sim0.32$ and the $z\sim0.44$ case are at the tail of the posterior distribution in the no prior case, resulting in a mild $\sim 1\sigma$ tension. This may be due to the limitation of our simplified halo model, or signal loss that is uncorrected for. A more detailed study into the implication of the $\Omega_{\rm HI}$ constraints is beyond the scope of this work.

\begin{figure*}
\centering
\includegraphics[width=0.45\textwidth]{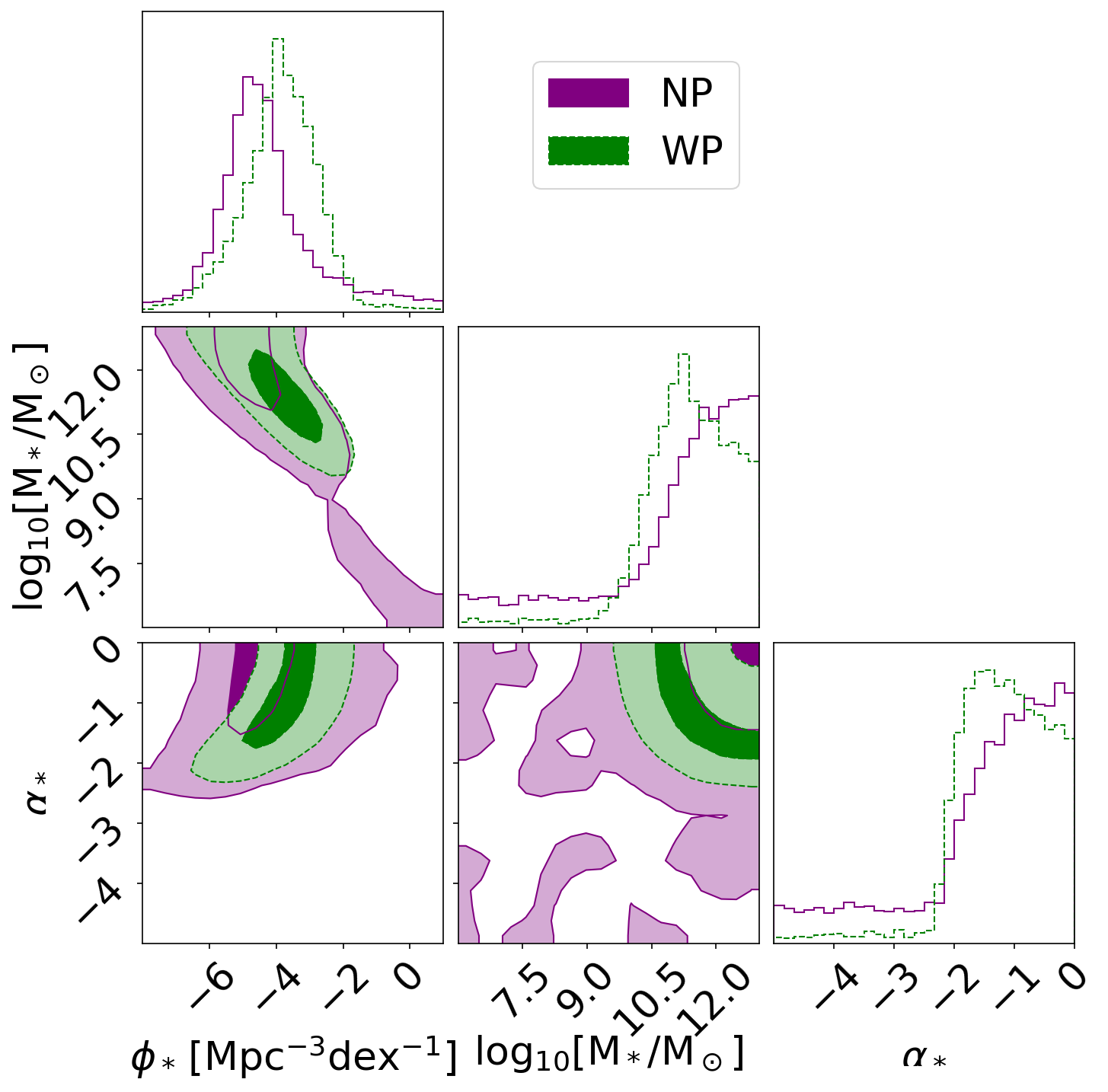}
\includegraphics[width=0.45\textwidth]{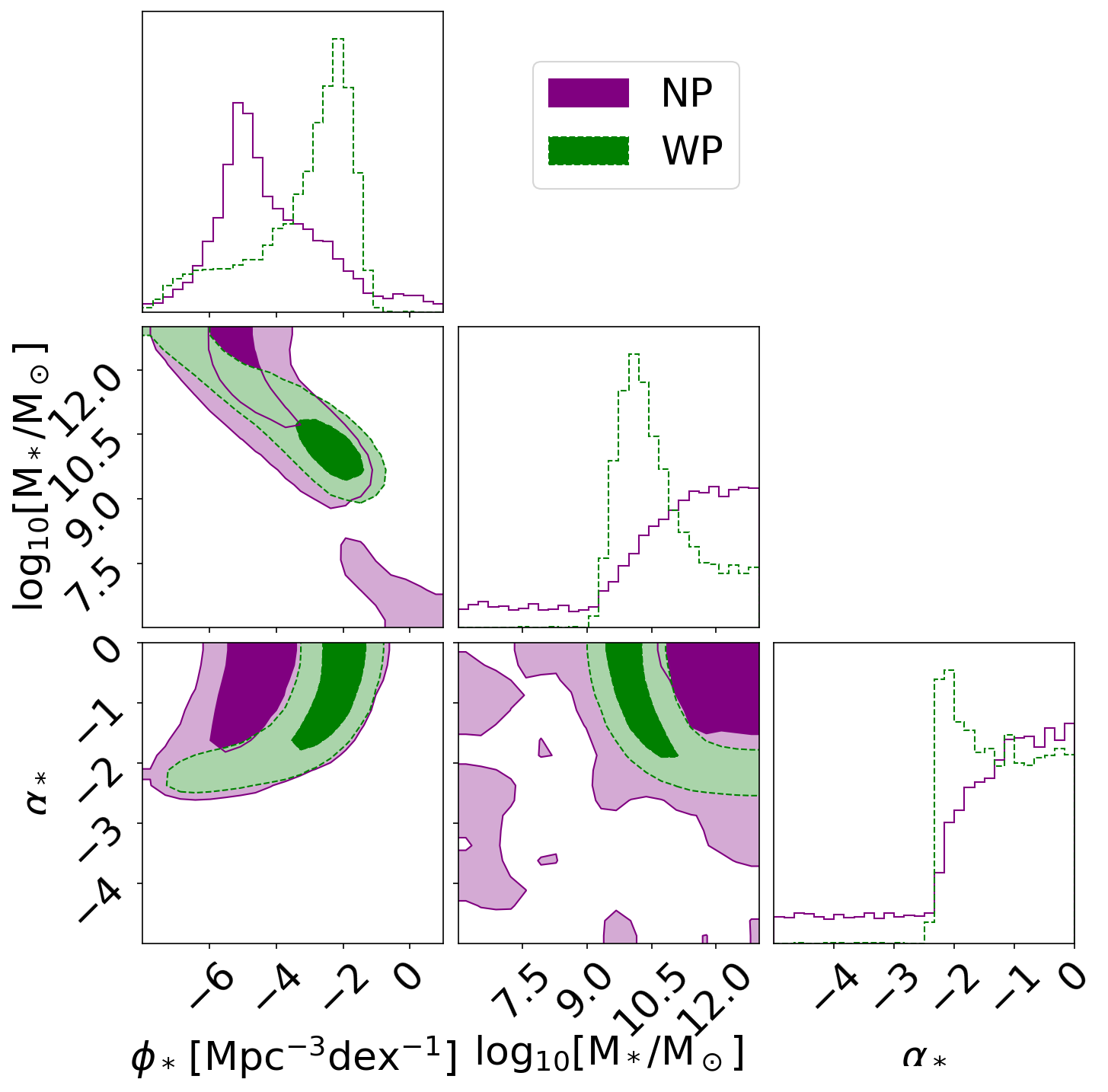}

\caption{{\bf Model fitting results for the HIMF parameters.} The 2-d contour plots show the posterior distribution of HIMF parameters. The darker region shows the $1\sigma$ confidence interval of the posterior and the lighter region shows the $2\sigma$ confidence interval. The upper lines of the titles of the 1-d histograms show the median and the $1\sigma$ confidence level for each parameter for the model fitting with the $\Omega_{\rm HI}$ prior (``WP''). The left panel shows the results for $z\sim 0.32$ and the right panel shows the results for $z\sim 0.44$.}
\label{fig:himf}
\end{figure*}

Using the posterior on $\Omega_{\rm HI}$ and the HIMF parameters $\{M_*, \alpha\}$, we can calculate the posterior for the other HIMF parameter $\phi_*$ to reconstruct the posterior for the HIMF. We shot the results in \autoref{fig:himf}. As expected, due to the HI power spectrum not sensitive in the number density of galaxies, the HIMF can not be fully constrained.
In particular, there is strong degeneracy between the knee mass $M_*$ and $\alpha$.

\begin{figure}
\centering
\includegraphics[width=\linewidth]{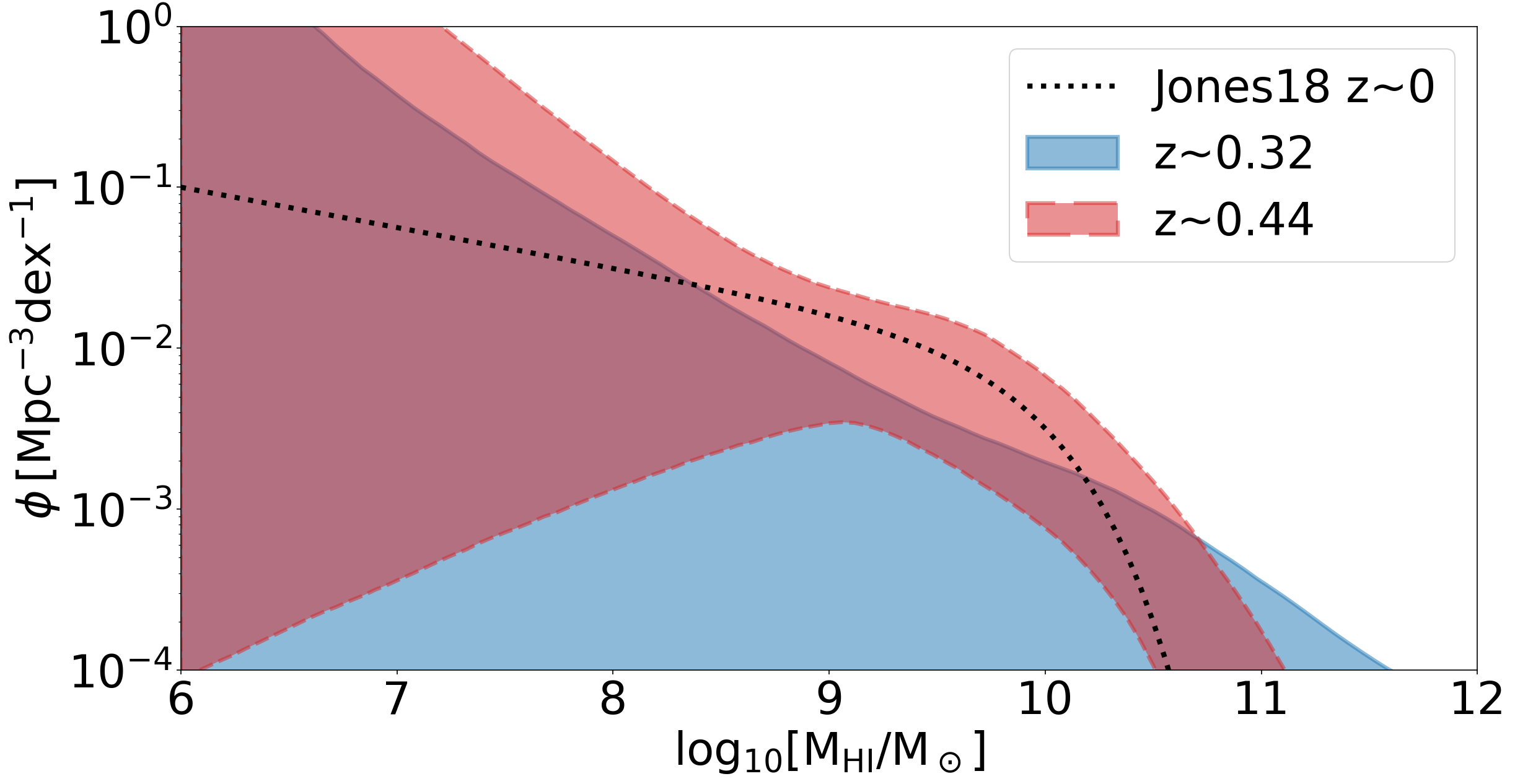}
\caption{The $1\sigma$ confidence interval for the reconstructed HIMF from the fitting. The blue region enclosed by the solid line shows the results for $z\sim 0.32$ and the red region enclosed by the dashed line shows the results for $z\sim 0.44$. For reference, the measured HIMF at $z\sim 0$ from \citet{Jones_2018} is also shown.}
\label{fig:himf2}
\end{figure}

Using the posterior of the HIMF parameters, we can reconstruct the posterior for the HIMF, which we show in \autoref{fig:himf2}.
For the $z\sim0.32$ bin, we find the fitting produces an upper limit on the HIMF.
Compared with the HIMF measured at $z\sim0$, the 68\% confidence interval suggests that at $M_{\rm HI}\sim 10^9\,M_\odot $, the number density of HI galaxies is smaller at $z\sim 0.32$.
For the $z\sim 0.44$ bin, the 68\% confidence interval produces constraints on the HIMF. Compared with the HIMF at $z\sim0$, we find that the HIMF at $z\sim 0.44$ may have a slightly higher distribution at the high mass end $M_{\rm HI}\gtrsim 10^{10.5}\,M_\odot$.

Our results show that using MeerKAT interferometric observations, we can constrain the velocity dispersion and the HI shot noise even with preliminary measurements with a relatively low signal-to-noise ratio. In the future with more data from the MeerKAT observations, intensity mapping measurements can be used to constrain the halo model of HI, providing a detailed picture of the galaxy evolution from $z\sim 0.0$ to $z\sim 1.0$.

\section{Summary}
In this letter, we use the MeerKAT observations of the DEEP2 field to derive the first-ever detection of the HI auto power spectrum at $z\sim 0.32$ and $z\sim 0.44$, respectively, measured with 220 frequency channels around either central redshift. The deep observations allow accurate calibration of the data with errors $\sim 10^{-5}$. We divide the visibility data into two data cubes from different timesamples (`odd' and `even') to minimise time-dependent systematics and remove noise bias, which allows us to measure the cross power spectrum of the cubes. We use the foreground avoidance technique, however, we detect the presence of weak broadband RFI which contaminates our observation window. We present two different approaches to detect and flag these systematics. The Baseline Flagging approach (BLF) focuses on the early detection and flagging of contaminated antenna pairs per scan, before proceeding to power spectrum calculations. In contrast, $uv$ Delay Flagging (UVDF) applies a thresholding criterion after gridding the visibility data, comparing the auto-power spectrum with the expected thermal noise for odd and even datasets separately. Both methods yield consistent results, as illustrated in \autoref{fig:1d_PS}, where the resulting power spectrum measurements in the $k_\parallel > 0.3 k_\perp$ region are binned into 1-d $k$-bins. The HI power spectrum at both redshifts is measured with high statistical significance of 3.2$\sigma$ and 3.5$\sigma$ for BLF and 5.9$\sigma$ and 9.18$\sigma$ for UVDF flagging, respectively. We also present a suite of tests to validate each step of our pipeline which includes simulating foreground scatter due to calibration errors, testing signal loss or bias due to flagging, inpainting and gridding with simulations, jackknife tests and cross-correlating the timeblocks between the two frequency sub-bands as a null test. 

The resulting HI power spectrum is then used to perform parameter inference. Using halo model formalism to calculate the HI power spectrum and including the effects of redshift space distortion, we find that the scales probed by our measurements are dominated by the combination of the HI shot noise and the velocity dispersion. Due to the fact that the power spectrum is measured in 1-d $k$-bins, losing the information of the $k_\parallel$ dependence, the parameters are degenerate with each other, leading to a divergence in the fitting. Imposing a $\Omega_{\rm HI}$ prior based on existing constraints, we find that preliminary constraints on the HI shot noise, the velocity dispersion, and the HI density profile can be made. 

Our results show that it is possible to isolate the HI signal from the strong foregrounds using the avoidance technique, however, they also demonstrate how weak, broadband RFI poses a risk to sensitive cosmological observations. This study highlights the need to monitor and control the RFI environment of the next generation of highly sensitive radio telescopes. MeerKAT surveys such as MIGHTEE \citep{MIGHTEE} and Laduma \citep{LADUMA} are currently undertaking deep observations on several fields suitable for cosmological analysis. If RFI contamination can be mitigated, they will allow to measure the small scale HI power spectrum to high precision and provide exquisite constraints on the HI mass function, the halo model and the nature of redshift space distortions on these scales.

\section*{Data availability}
The uncalibrated visibility data used in this work are publicly available under the Proposal ID: SCI-20180426-TM-01 in SARAO archive (\url{https://archive.sarao.ac.za}).

\section*{Code availability}
The initial processing and calibration of the MeerKAT data was carried out using the processMeerKAT pipeline developed at the IDIA and available at \url{https://idia-pipelines.github.io/docs/processMeerKAT}. The self~calibrations were performed with standard CASA tasks: \textit{tclean}, \textit{gaincal} and \textit{applycal} (\url{https://casa.nrao.edu/casa_obtaining.shtml}). Further codes are available on a reasonable request to the authors.   

\section*{Acknowledgements}
We are thankful to the anonymous referees for their careful and constructive reports, which led to a number of useful tests and clarifications and significantly improved the manuscript. MGS and SP acknowledge support from the South African Radio Astronomy Observatory and National Research Foundation (Grant No. 84156). SP acknowledges support from the Science and Technology Facilities Council (STFC) through the Consolidated Grant ST/X001229/1 at the Jodrell Bank Centre for Astrophysics, University of Manchester. LW is a UK Research and Innovation Future Leaders Fellow [grant MR/V026437/1]. We acknowledge the use of the Ilifu cloud computing facility - \url{www.ilifu.ac.za}, through the Inter-University Institute for Data Intensive Astronomy (IDIA). We thank Keith Grainge, Tom Mauch and Jordan D. Collier for useful discussions on the MeerKAT data used in this work. We also thank the developers of open-source Python libraries NumPy \citep{Numpy}, SciPy \citep{Scipy}, emcee \citep{Foreman-Mackey_2012}, corner \citep{corner}  and Matplotlib \citep{Matplotlib}. 
The MeerKAT telescope is operated by the South African Radio Astronomy Observatory, 
which is a facility of the National Research Foundation, an agency of the Department 
of Science and Innovation.

\appendix
\section{Polarization Leakage}
\begin{figure}[!h]
    \centering
    \includegraphics[width=1.0\linewidth]{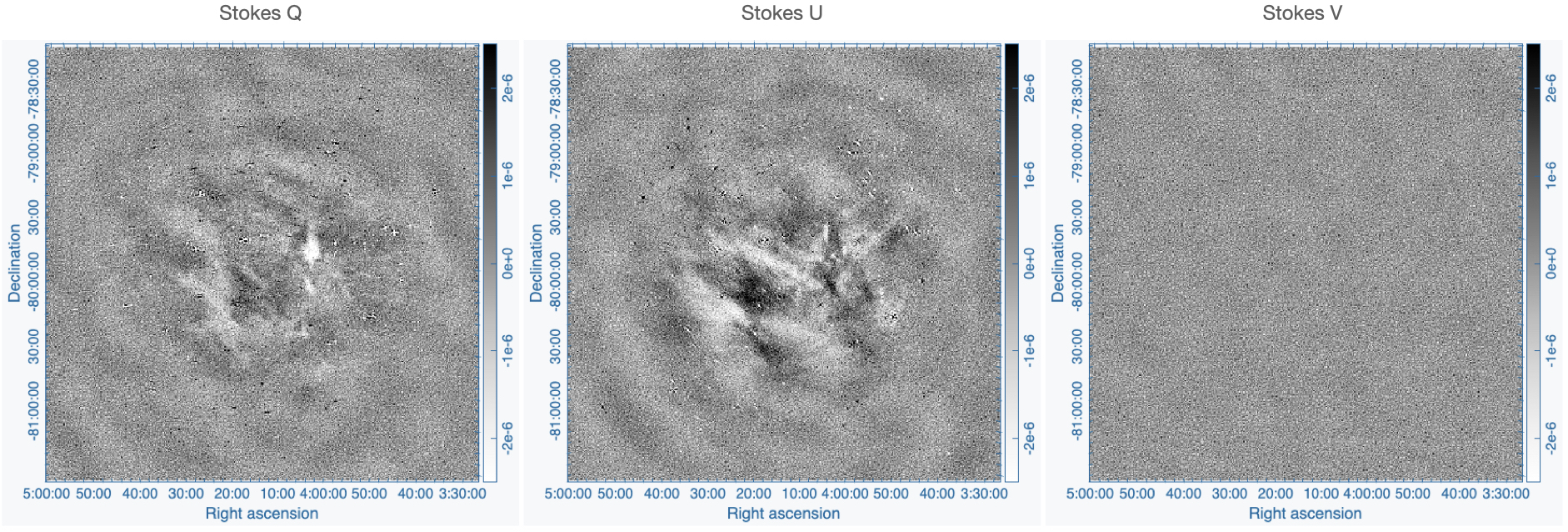}
    \caption{Frequency-averaged images of the Stokes Q, U and V modes at $1077.5$ MHz, generated from 1 dataset of $12.75$ hours duration.}
    \label{fig:Stokes_QUV}
\end{figure}

\begin{figure}[!h]
    \centering
    \includegraphics[width=1.0\linewidth]{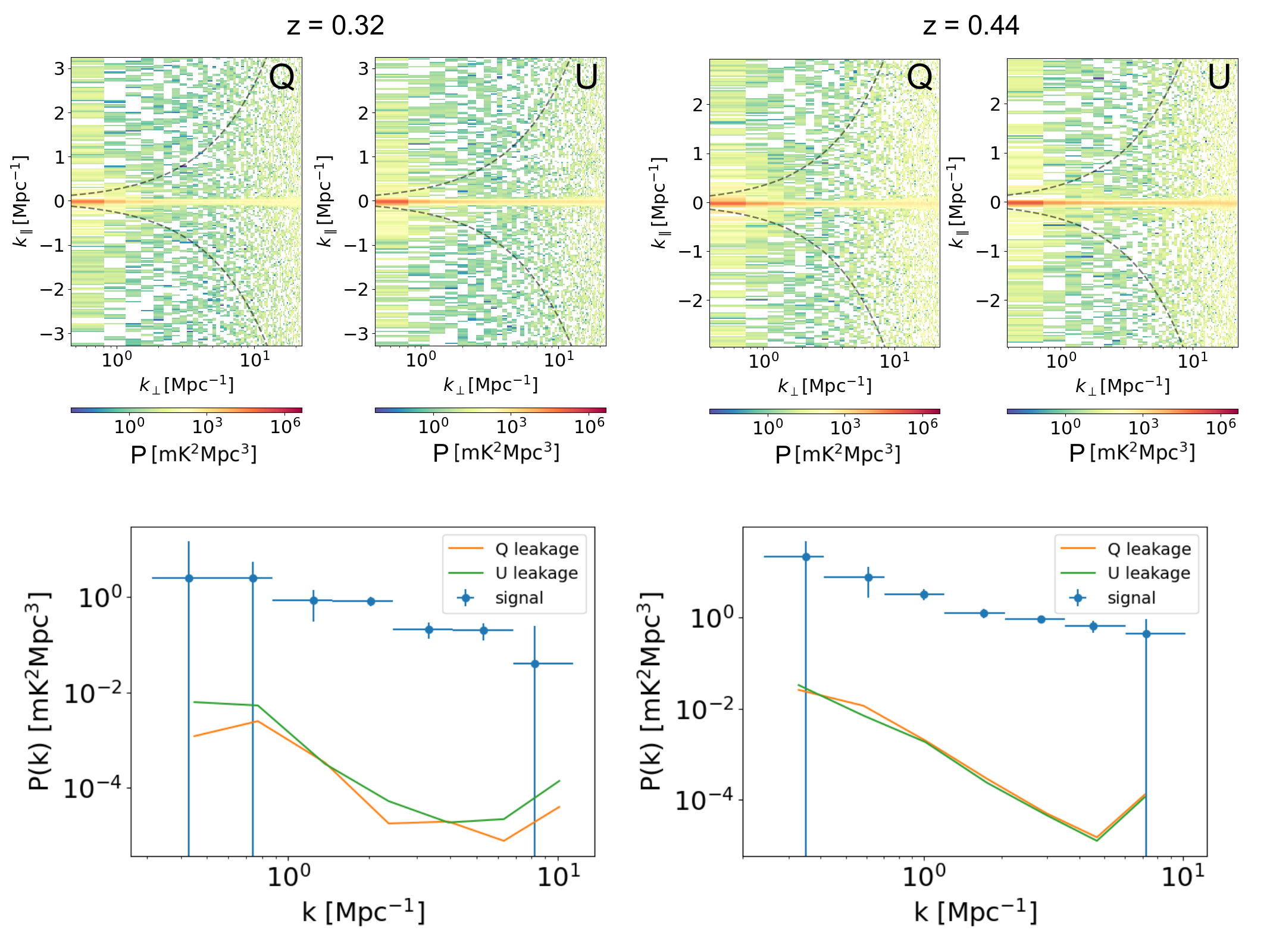}
    \caption{The top panel shows the Stokes Q and U power spectrum (2-d) for $z=0.32$ and $z=0.44$ cases. The bottom panel shows the expected leakage from Stokes Q/U to I assuming 1\% leakage in the visibilities. The measured signal in Stokes I with UVDF is also shown for comparison (signal).}
    \label{fig:QU_leakage}
\end{figure}

There is no dedicated polarization calibrator source observed in this dataset. However, MeerKAT performs delay calibrations every night and calculates cross-polarisation solutions using noise diode injections, which are stored in the metadata. So, Stokes I and Q can be calibrated using the observed unpolarised calibrators (which also measures some of the leakage terms), while the polarization tables are used to calibrate Stokes U and V\footnote{https://archive-gw-1.kat.ac.za/public/meerkat/MeerKAT-L-band-Polarimetric-Calibration.pdf}. There are still some leakage terms in stokes V, to first order in U and second order in I. This is what we observe in Stokes V on short baselines. However, since we only use the long baselines of Stokes V to normalize our model noise power spectrum, this does not pose an issue. 
In terms of the possible leakage of stokes Q/U into I, this should be below our signal level.
The on axis leakage terms from MeerKAT have been measured to be at the level of 1\% or less in the visibilities\footnote{https://skaafrica.atlassian.net/wiki/spaces/ESDKB/pages/1484128294/Dynamic+range+considerations}. Thus, even if we assume that all the measured Stokes Q is real, 1\% of it would still produce a power spectrum that remains below our signal in the target window. In \autoref{fig:Stokes_QUV}, we present continuum images for Stokes Q, U, and V from one dataset. The images for Stokes Q and U display diffuse structures on a large scale, which are likely to be polarized synchrotron emissions from the Milky Way. On the other hand, the Stokes V image is primarily noise-like, indicating that the polarization calibration has performed reasonably well. \autoref{fig:QU_leakage} shows the Stokes Q and U power spectrum in 2-d, along with expected leakage from Q/U to I in the 1-d power spectrum.

\section{Primary Beam model}
\label{sec:PrimaryBeam}
\begin{figure}[!h]
    \centering
    \includegraphics[width=1.0\linewidth]{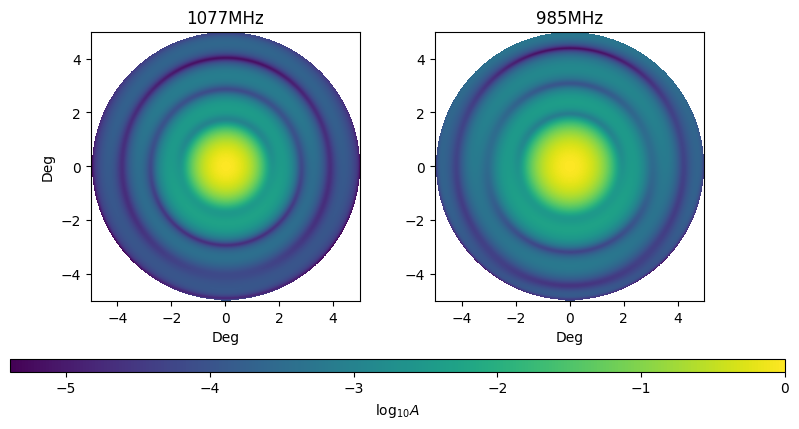}
    \caption{Primary beam in Stokes I using eidos based on the ``astro-holographic'' (AH) observations of the MeerKAT L-band beam. The beam description is only accurate up to 5 deg.}
    \label{fig:beam_2d}
\end{figure}

The primary beam model for our power spectrum calculation is generated using the \href{https://github.com/ratt-ru/eidos}{eidos} package, which is based on ``astro-holographic'' (AH) observations of the MeerKAT L-band beam \citep{2021MNRAS.502.2970A}. In Fig. \ref{fig:beam_2d}, we show the primary beam model for the frequency ranges we choose at the central frequencies 1077MHz and 985MHz. The MeerKAT L-band beam is $\sim 1$deg and follows a Cosine beam pattern (see Fig. 2 of \href{https://arxiv.org/abs/2011.10815}{Matshawule et al.}). The eidos beam is then used to calculate the beam area for conversion between flux and brightness temperature. Note that while we only model the primary beam down to -40 dB, it is sufficient for several reasons. First, the calibration is done by tracking calibrator sources so there is no requirement to model the beam to large angular distances. Second, the MeerKAT dishes have a smaller field of view (FoV) compared to EoR instruments, and we choose the region above the horizon value in the cylindrical $k$-space for our measurements.

\section{On the sky curvature}
We consider the sky curvature to be negligible for the DEEP2 data. First, the primary beam area is small for the MeerKAT dishes, with the FoV $\sim 1$deg (and power-squared beam being 0.5 deg) and the beam extends to 5deg across, where flat-sky approximation holds well. As shown by \citet{2023MNRAS.518.2971C}, simulations of the HI signal for MeerKAT observations without $w$-projection, while retaining the $w$-kernel in the simulation, demonstrate that the HI power spectrum can be successfully recovered. Note that  $w$-projection kernels are deconvolved when constructing the foreground sky model for simulations and for self-calibrations.

\section{RFI flagging and inpainting}
\label{sec:inpainting}
The RFI flagging is done using the standard \href{https://idia-pipelines.github.io/docs/processMeerKAT}{\textsc{processMeerKAT}} pipeline (discussed in \autoref{Data_calibration}). Additionally, we exclude baselines with more than 20\% of the frequency channels flagged to further mask potential wide-band RFI. The 20\% threshold is calculated for each individual sub-band, not the overall L-band frequency range. For the $z=0.32$ redshift bin, there is severe contamination at around $1090$ MHz with more than 50\% of the channels flagged. The effect of the flagging can be seen in \autoref{fig:ps2dfg} in terms of foreground scatter.

\begin{figure}[!h]
    \centering
    \includegraphics[width=\linewidth]{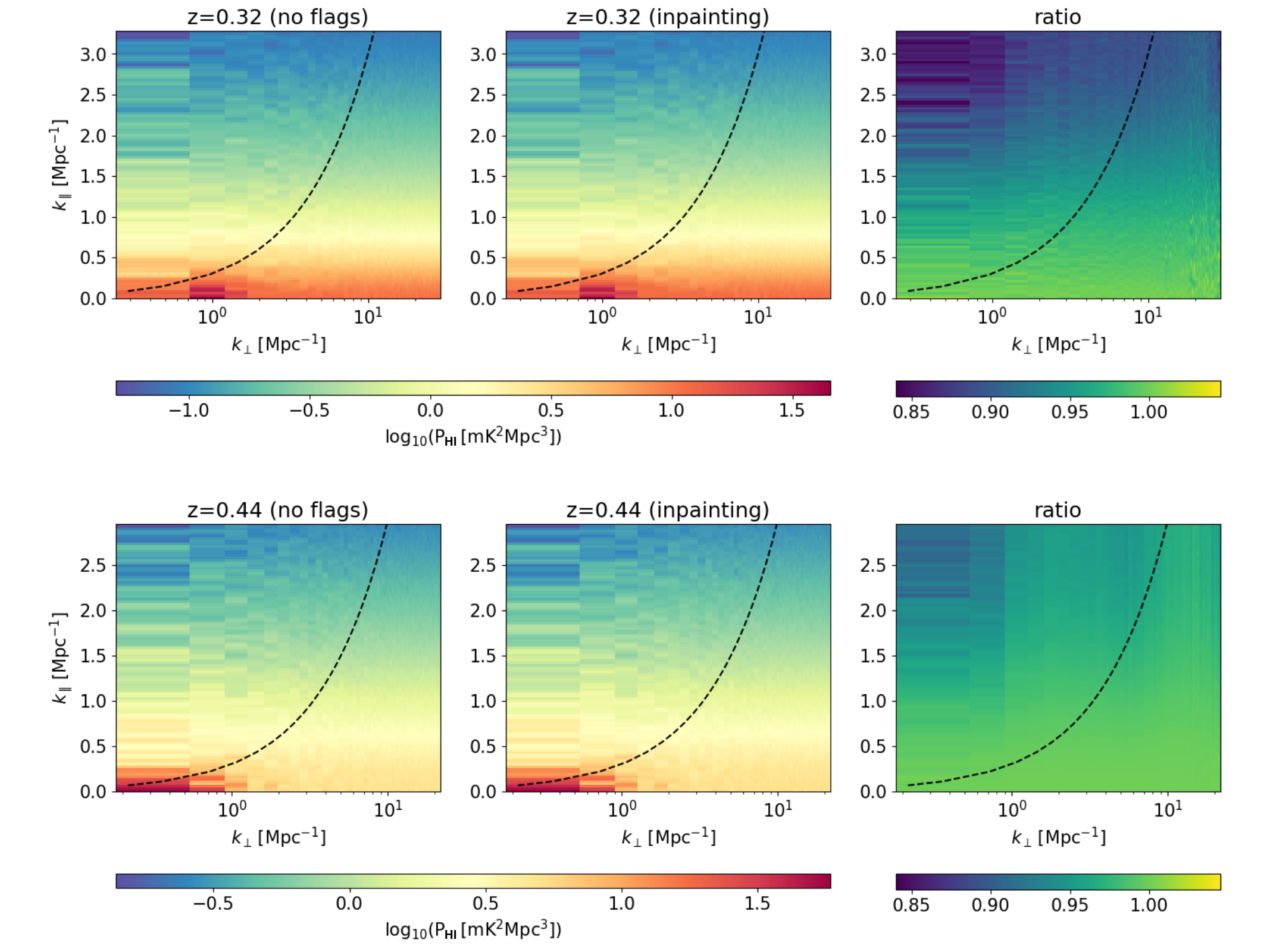}
    \caption{Top panels: The cylindrical HI power spectrum from simulated HI visibilities of one realization at $z=0.32$. The left panel shows the HI power spectrum without any flagging and inpainting. The central panel shows the power spectrum with the exact same flags as the data with the subsequent inpainting. The right panel shows the ratio between the central panel and the left panel. Bottom panels: The same as the top panels for $z=0.44$. It shows that inpainting causes signal loss in the HI power spectrum at high $k_\parallel$.}
    \label{fig:hicypower}
\end{figure}

As discussed before, the inpainting of flagged channels is done to reduce foreground scatter from the delay transform. This is applied by filling the flagged channels with the nearest neighbours. While this can lead to non-negligible numerical effects if it is done for a large fraction of the frequency channels, we keep it under control by excluding baselines with more than 20\% of frequency channels flagged. The impact of our flagging and inpainting on the foregrounds has already been discussed (\autoref{fig:ps2dfg}). In \autoref{fig:hicypower}, we show the impact of inpainting on the cylindrical power spectrum from the simulated HI visibilities. The inpainting procedure does lead to some signal loss in the HI power spectrum, in particular in the low $k_\perp$, high $k_\parallel$ region. This is more obvious for the $z=0.32$ bin where there is more flagging. Averaging the powers into the 1-d power spectrum as shown in \autoref{fig:ps1dhisim_alt}, we find that the signal loss caused by flagging/inpainting is negligible compared to the measurement error.

\section{Robustness of the Detection to Horizon Buffer Size}
\label{sec:buffer_test}
While our fiducial analysis adopts a horizon buffer of $k_{\parallel} > 0.3 k_{\perp}$ to mitigate foreground spillage and instrumental chromaticity, it is crucial to verify that the reported detection is not driven by residual systematics concentrating near the horizon boundary (\autoref{fig:2d_PS_1}). To test the robustness of our signal, we re-computed the power spectra for both redshift bins using a stricter foreground avoidance criterion, increasing the buffer size to $k_{\parallel} > 0.35 k_{\perp}$. If the detected signal were primarily due to foreground leakage or horizon-delay systematics, we would expect a significant reduction in power as the exclusion zone is expanded.

\begin{figure*}[h]
    \centering
    \includegraphics[width=\textwidth]{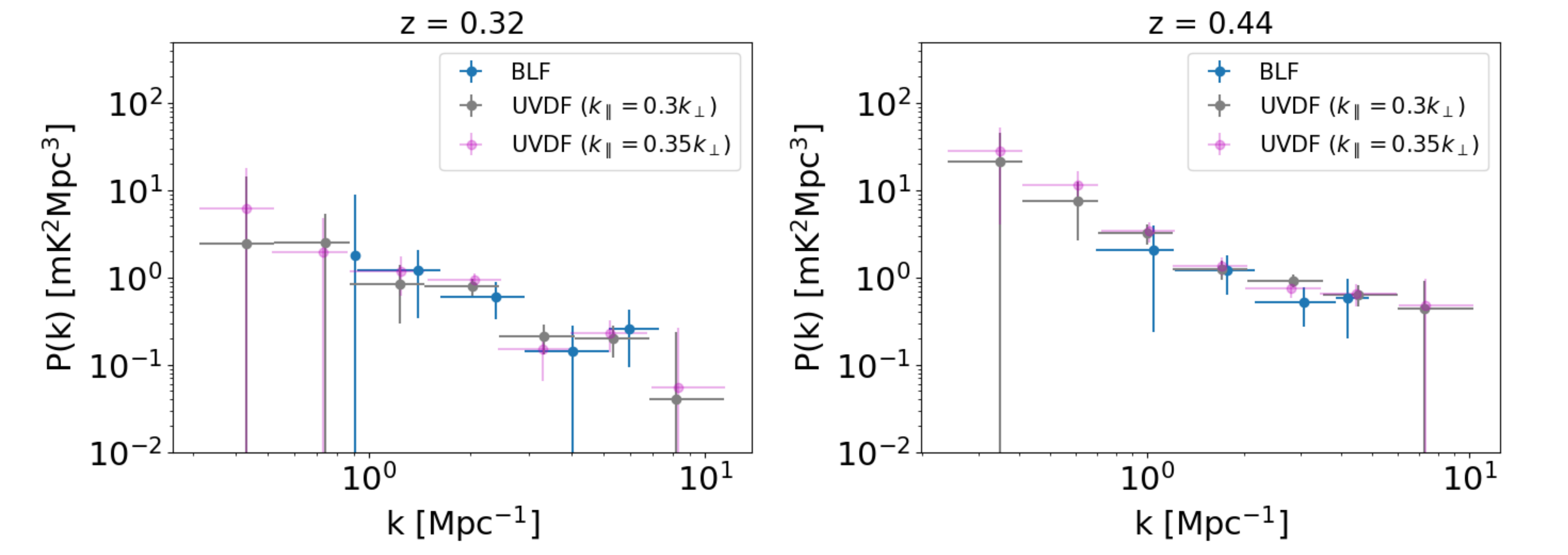}
    \caption{{\bf Robustness of the HI power spectrum to the horizon buffer size.} We compare the results from our fiducial UVDF analysis using the standard horizon cut ($k_{\parallel} \ge 0.3 k_{\perp}$) against a stricter cut ($k_{\parallel} \ge 0.35 k_{\perp}$). The results from the BLF method are shown for reference. The consistency of the signal amplitude between the two UVDF cases demonstrates that the detection is robust and is not dominated by foreground leakage or systematics near the horizon wedge.}
    \label{fig:buffer_test}
\end{figure*}

The results of this test are presented in \autoref{fig:buffer_test}. We observe that the power spectra derived using the stricter $0.35 k_{\perp}$ cut are statistically consistent with our fiducial results across the measured scales for both redshift bins. Specifically, the amplitude of the detection does not collapse to the noise level or the lower amplitude. Quantitatively, we compare the fiducial $k_\parallel > 0.3k_\perp$ UVDF measurement with the stricter $k_\parallel > 0.35k_\perp$ measurement using only the robust $k$-bins, excluding the first two low-$k$ bins that are more susceptible to residual systematics and are also excluded from the fiducial detection significance quoted in this work. Over these five $k$-bins, we find $\chi^2=0.81$ ($\chi^2_\nu=0.16$) at $z=0.32$ and $\chi^2=0.68$ ($\chi^2_\nu=0.14$) at $z=0.44$, using a diagonal error estimate and neglecting covariance between the two highly overlapping measurements. The stricter-cut spectra also remain inconsistent with the null hypothesis at $6.2\sigma$ and $8.1\sigma$ for $z=0.32$ and $z=0.44$, respectively.

This stability indicates that the signal is distributed throughout the clean $k$-space window rather than being confined to the wedge boundary, thereby ruling out foreground leakage or residual systematics as the primary source of the detected power.

\section{Impact of BLF on $uv$-Sampling and Sensitivity}
\label{sec:blf_sensitivity}
In \autoref{sec:1d_ps_comparison}, we noted a significant difference in detection significance between UVDF and BLF methods. To investigate whether this discrepancy arises from the removal of a spurious signal or a loss of sensitivity, we analyzed the spatial distribution of flagged data in the $uv$-plane.

\begin{figure}[h]
    \centering
    \includegraphics[width=1.0\linewidth]{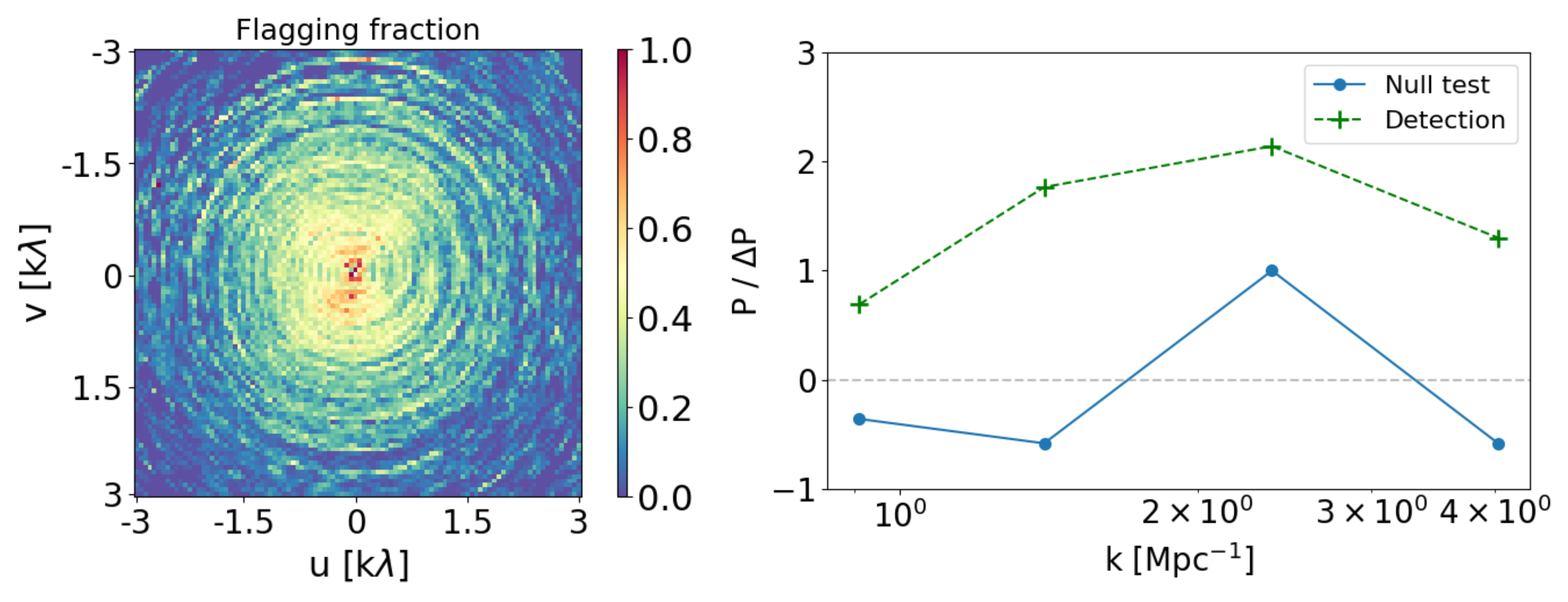}
    \caption{\textbf{Left:} Fraction of visibilities flagged by the BLF method. The color scale represents the fraction of data removed in each $uv$-cell, ranging from 0 (all data retained) to 1 (100\% data loss). The flagging is highly non-uniform, with the central region suffering severe data loss. Since the large-scale HI signal is predominantly contained in these short baselines, this aggressive flagging drastically reduces the sensitivity, inflating the thermal noise and reducing the detection significance. \textbf{Right:} Null test using the BLF method. The ratio of the cross-power to its $1\sigma$ error ($P/\Delta P$) is shown. The `Null test' (cross-correlating different $z$-bins) is consistent with zero, confirming that BLF removes the broadband systematics. The `Detection' significance is lower than the UVDF case, which is consistent with the inflation of the thermal noise due to the severe data loss shown in the left plot.}
    \label{fig:blf_fraction}
\end{figure}

\autoref{fig:blf_fraction} (left) shows the fraction of time samples flagged by the BLF algorithm as a function of $uv$-coordinates. The flagging exhibits a strong radial dependence: while long baselines remain relatively unflagged, the short baselines ($|u| \lesssim 1000 \lambda$)—which correspond to the largest angular scales—are heavily impacted. In this central high-sensitivity core, we find that the BLF method discards an average of $>50\%$ of the available data. This extensive data loss has a critical impact on the power spectrum estimation. The thermal noise variance of the power spectrum, $P(k)$, scales inversely with the number of independent modes ($N_m$) contributing to each $k$-bin. By systematically hollowing out the dense core of the $uv$-plane, BLF removes the baselines that dominate the sensitivity to the cosmological signal. Consequently, the reduction in detection significance observed in the BLF case is driven primarily by a substantial increase in the thermal noise floor, rather than a suppression of the signal amplitude itself.

We also perform the same Null Test (cross-correlating the two redshift bins) on the BLF-flagged data. The results are shown in \autoref{fig:blf_fraction} (right) which is consistent with zero. This mirrors the behavior of the UVDF method (\autoref{null_test}), proving that both flagging strategies successfully isolate and remove the broadband systematics.

\section{Expected level of HI signal}
\label{sec:model}
In this section, we briefly discuss the expected amplitude of the HI power spectrum for our measurements shown in \autoref{1d_ps}. While typically we can use hydrodynamical simulations to inform us on the amplitude of the HI power, it is hard to do so for our measurements. 

For scales $k>1.0\,{\rm Mpc^{-1}}$, the HI signal is dominated by the physics inside the dark matter halos and cannot be modelled by the cosmological perturbation theory. While it may be empirically described by the halo model, it is unclear how to model the redshift space distortion effects at these scales in the context of the halo model.
From the simulation side, \citet{Villaescusa-Navarro_2018} reports that the redshift space distortions found in the simulation are not in agreement with the description of the finger-of-god effects in halo model (see Section 16). \citet{2020ApJ...895...34Z} claims refined modelling of the RSD on the shot noise can be used to improve the accuracy of the modelling. But none of these works addresses the small scales we are probing, and they compare power spectrum monopoles, which are not equivalent to our 1-d power spectra due to the foreground avoidance.

Simulations of different approaches also start to have significant discrepancies down to small scales. For example, in \citet{Villaescusa-Navarro_2018} the shot noise is modelled as the $k\rightarrow0$ limit of the one-halo term (see Section 15). In \citet{Spinelli_2019}, it is modelled as the square sum of the HI galaxy masses, which is lower than the 1-halo term (see Fig. 10). In short, we expect that a halo model framework with the separation of central and satellite galaxies including FoG effects can be used to constrain the measurements, which is what we use in the paper for a proof of concept.

From the observation side, it would still be possible to estimate the signal level if we know the HIMF and the velocities of the HI galaxies. The HIMF gives us the shot noise in the comoving space where the velocities give us the FoG effects. Future measurements using stacked HI galaxies will serve as a tool to validate the signal level we detect in the power spectrum. 

Based on the simulations as well as measurements of HIMF in the local Universe \citep{Jones_2018}, the HI power spectrum should be $\sim 10\,{\rm mK^2Mpc^3}$ in comoving space. The attenuation from the velocity dispersion, typically at $\sigma_v \sim 100\,{km/s}$ (e.g. \citealt{2021MNRAS.508.1195P}), should lower the signal to $\lesssim 1\,{\rm mK^2Mpc^3}$ in redshift space at small scales. The strong degeneracy and lack of understanding of the shot noise as well as the velocity dispersion means that, even if we fix the halo model and only vary these two parameters in plausible ranges, the power spectrum covers several orders of magnitude.

To illustrate this, we assume two different sets of values for the shot noise and the velocity dispersion, and use the range in between as the expected signal level as shown in \autoref{1d_ps}. In particular, we set the upper limit as $P_{\rm SN} = 40\,{\rm mK^2Mpc^3}$ and $\sigma_v = 50\,$km/s, and the lower limit as $P_{\rm SN} = 1\,{\rm mK^2Mpc^3}$ and $\sigma_v = 500\,$km/s. Note that we are only varying the shot noise and velocity dispersion parameters, which dominate on small scales. The power spectrum amplitude we measured is completely in line with the expectations.


\bibliography{references}{}
\bibliographystyle{aasjournal}

\end{document}